\documentclass[11pt]{article}
\usepackage{amsmath}
\usepackage{amssymb,graphicx,subfigure}
\usepackage{cite}
\begin{document}

\title{\bf Gluonic Structures of Tetraquarks}
\author{Jin Dong Kim\footnote{E-mail : ph8858@hanmail.net}\\
{\sl {\small{ Department of Physics and Technology,}}}\\
{\sl {\small{Chonbuk National University, Chonju 561-756, Korea}}}\\
\\
Su Kyoung Lee$^{a}$\footnote{E-mail : sklee92@chonbuk.ac.kr} and Jong B. Choi$^{a,b}$\footnote{E-mail : jbchoi@chonbuk.ac.kr}\\
{\sl {\small{$^a$Institute of Proton Accelerator, Chonbuk National
University, Chonju 561-756, Korea}}}\\
{\sl {\small{$^b$Department of Physics, North Carolina State
University, Raleigh, NC 27695}}} }

\date{}
\maketitle \vspace{-9mm}
\begin{abstract}
We formulate quantitative flux-tube overlap functions to account for
the nonperturbative gluonic behaviors. In order to deduce systematic
functional form, we introduce connection amplitude which can be
defined between two boundary points occupied by quark or antiquark.
With the deduced flux-tube overlap function, we can figure out long
range gluonic structures of various quark combinations. In this
paper, we report our calculated results for tetraquarks, $q^2
\bar{q}^2$, by considering two different configurations of boundary
points for possible explanations of $X(3872)$.
\end{abstract}

%1절
%
\section{Introduction}
\label{one}\vspace{-3.5mm}
 The quantum numbers of hadrons such as mesons and baryons are
usually described by combining those of valence quarks. However,
from the dynamical viewpoint, the binding forces are generated by
gluons. These gluons not only generate binding forces, but also
constitute important parts of hadrons affecting the long range
structures. For the simplest meson system, mass spectra can be
estimated by considering only one gluon exchange diagrams\cite{R1},
whereas the next complicated baryon system cannot be easily analyzed
without introducing gluonic degrees of freedom\cite{R2}. Even for
the meson system, strong decay processes cannot be accounted
appropriately without considering gluonic effects\cite{R3}. These
gluonic effects are in the realm of nonperturbative interactions,
which can only be described by phenomenological models such as
flux-tube model.\par
 Gluonic flux-tube picture was introduced to put the gluonic degrees of
freedom explicitly into the calculations of baryon spectra\cite{R2},
and later, quantitative approach was made to predict strong decays
of mesons\cite{R3}. Strong decays of hadrons are induced by quark
pair creations which are closely related to the gluon densities or
gluonic energies. Since the position of created quark pair affects
the final states of strong decay, it is important to check the
gluonic profiles of initial state. For mesons, it was firstly
attempted to introduce cigar-shape profiles, which were changed into
more generalized forms including the case of spherical
shape\cite{R4}. The consideration of spherical shape of gluonic
flux-tube is motivated by the calculational convenience in treating
quark wave functions. Strong decay amplitudes are dependent on the
product of quark wave functions and gluonic flux-tube overlap
functions, and therefore it is convenient to use the same form of
function for the quarks and the gluons if possible. One possible
choice is the harmonic wave functions which can be used safely for
quark motions and the Gaussian form can be applied to describe
gluonic behaviors. Because the Gaussian form is a function of
radius, the resulting gluonic structure becomes that of spherical
shape. However, it is quite arbitrary whether we choose the cigar
shape or the spherical shape. Each choice has its merits and
defects, and there existed no consensus on how to fix the form of
flux-tube overlap function.\par
 In order to deduce systematically the form of flux-tube overlap function, we need to analyze the
structures of flux-tubes and to introduce some quantitative function
that fits to the description of those structures. We can start the
analysis by classifying the flux-tubes resulting in the natural
consideration of various structures such as tetraquarks,
pentaquarks, hexaquarks, and so on. The relations between these
structures can be established by taking account of quark pair
creations and quark pair annihilations with the effects of flux-tube
breakings and connections respectively. The related flux-tubes can
be shown to form topological spaces\cite{R5}, and we can define
physical amplitudes on the closed topological spaces\cite{R6}. For
the physical amplitude, we are seeking for something that can be
used to describe gluonic states. Since the long range gluonic states
are described by flux-tubes, the connectedness through given
flux-tube between two quarks can be chosen to define appropriate
measure that can be used to estimate gluonic states. In this way, a
systematic definition of flux-tube overlap function can be made and
applied easily to various situations such as tetraquarks and
pentaquarks in addition to the ordinary mesons and baryons\cite{R7}.\par
 The tetraquarks and pentaquarks have peculiar positions between ordinary
hadrons and nuclei. For mesons and baryons, there exist so nice
models to account for their mass spectra and decay processes that
their quark compositions as $q\bar{q}$ and $qqq$ can never be
replaced by other ones. On the other hand, stable nuclei can be
explained more appropriately by the bound states of nucleons rather
than by quarks and gluons. Even the simplest nucleus, deuteron,
cannot be easily described by quarks and gluons. Therefore it is
quite important to check whether we can extend the descriptions by
quarks and gluons to more complicated systems. In this respect,
tetraquarks and pentaquarks are good candidates to be studied by
using quarks and gluons. Recent discoveries of $X(3872)$\cite{R8}
and $\Theta^+(1540)$\cite{R9} stimulated the concerns about the
structures of these states. However, as is well-known, the
pentaquarks were predicted in the context of chiral soliton model
rather than in the picture of five quarks bound by gluons\cite{R10}.
The problems of 4 or 5 bodies are quite difficult to treat, and
moreover, the non-linear gluonic interactions have to be considered
via new method such as flux-tubes.\par
 In this paper, we will devise a new method to set up flux-tube overlap functions and apply
them to tetraquark configurations. Since we don't know the positions
of quarks, we will consider two cases, one in the form of
tetrahedron and the other in rectangular shape. We can calculate
gluonic structures for each cases which can be compared
quantitatively resulting in the selection of lowest energy
configuration. In section \ref{two}, we will give general formalism,
and calculated results are given in section \ref{three}. The final
section is devoted to discussions and further problems.

%%%%%%%%%%%%%%%%%%%%%%(2 절)%%%%%%%%%%%%%%%%%%%%%%%%%%%%%%%%%%%%%%%%%%%%%%%
\section{Flux-tube Formalism}
\label{two} \vspace{-3.5mm}
 The flux-tube picture has been introduced to account for the long range
behaviors of gluons. Here the long range means that of near 1 fermi,
and for smaller region than 0.1 fermi we can apply perturbative
methods to calculate physical amplitudes. Therefore flux-tube
formalism is useful for the ranges between 0.1 fm and 1 fm. Since
ordinary hadrons are estimated to have the extensions of 1 fm,
flux-tubes can be applied at least to the descriptions of mesons and
baryons. In addition, we can generalize the structures of flux-tubes
to include many quark systems such as tetraquarks, pentaquarks,
hexaquarks\cite{R11}, and so on. These generalizations are closely
connected to the classification scheme of flux-tubes.\par
\begin{figure}[]
\centering\includegraphics[height=10cm,width=10cm  ]{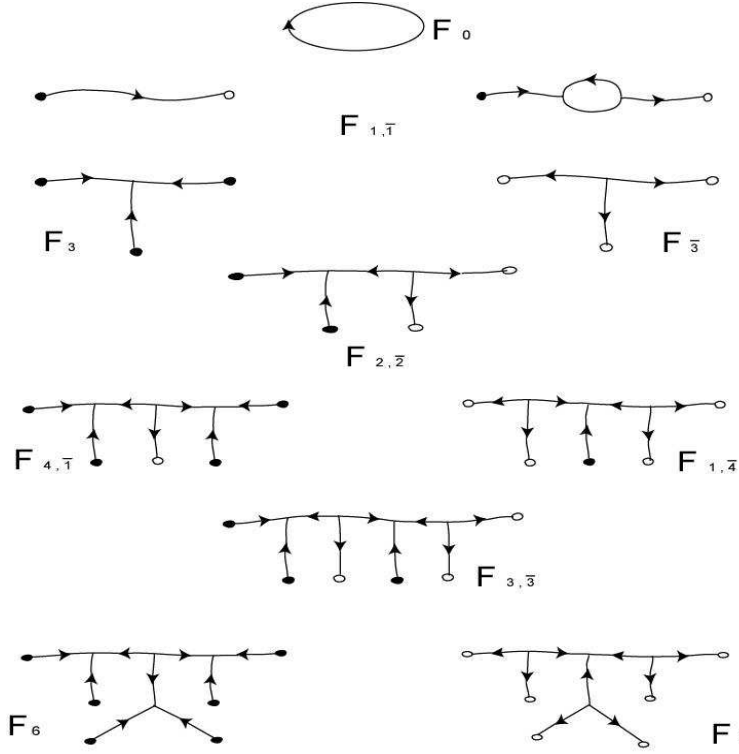}
\vspace{-3mm} \caption{\small{The classified flux-tubes.
$F_{2,\bar{2}}$ represent tetraquarks and $F_{4,\bar{1}}$ and
$F_{1,\bar{4}}$ represent pentaquarks.}} \label{fig1}
\end{figure}

 In order to classify flux-tubes, we need to count the number of boundary
points on which quarks or antiquarks sit. We can take the flux-tubes
as starting from quark boundaries and ending at antiquark
boundaries. Then we can represent the set of flux-tubes with $a$
quarks and $b$ antiquarks sitting at boundaries as $F_{a,\bar{b}}$.
Mesons are represented by $F_{1,\bar{1}}$, and baryons correspond to
$F_{3,\bar{0}}$ with $F_{0,\bar{3}}$ that of antibaryons. In
general, we can omit the number 0 without confusion so that $F_3$
represents baryon flux-tubes and $F_{\bar{3}}$ corresponding to
antibaryons. Exceptionally $F_{0,\bar{0}}$ can be reduced to $F_{0}$
which represents the flux-tubes of glueballs. In this notation, only
different topological configurations are distinguished and therefore
different energy states with the same topology belong to the same
set of flux-tubes. The classified flux-tubes are shown in
Fig.\ref{fig1} where we can find naturally the structures of
tetraquarks, pentaquarks, hexaquarks, and so on.

%%%%%%%%%%%%%%%%%%%%%%%%(figure 1)%%%%%%%%%%%%%%%%%%%%%%%%%%%%%%%%%%%
 Now let's consider relationships between flux-tubes. The topological structure
of a given flux-tube can be changed only when the number of boundary
points is changed. The changes of boundary points are induced by
quark pair creations and quark pair annihilations. If one quark pair
is created in a flux-tube, the tube will be divided into two, and
conversely two flux-tubes can be united into one through quark pair
annihilation. The changes of division and union can be combined to
establish relationships between flux-tubes. The most general
relationship turns out to be the construction of topological spaces
of flux-tubes. For the construction of topological spaces, we need
to write down the following assumptions :
\renewcommand{\labelenumi}{\theenumi.}
\begin{enumerate}
   \item  Open sets are stable flux-tubes.\vspace{-3.5mm}
   \item  The union of stable flux-tubes becomes a stable flux-tube.\vspace{-3.5mm}
   \item  The intersection between a connected stable flux-tube and disconnected
      stable flux-tubes is the reverse operation of the union.
\end{enumerate}

These assumptions are based on the correspondence between flux-tubes
and topological open sets, and the operations of union and
intersection are quite natural with respect to the changes of
flux-tubes. If we find out closed sets under the operations of union
and intersection, we can classify flux-tubes into different
topological spaces. In fact, this classification can be made by
counting the numbers of incoming and outgoing 3-junctions in a given
closed set.\par
 For examples, let's consider first the flux-tubes $F_{1,\bar{1}}$ corresponding to
simple quarkonium mesons. If quark pair creations are repeated
disconnecting the flux-tubes, we get in general n quarkonium meson
states represented by $F^{n}_{1,\bar{1}}$. The inverse union process
does not change the situation, so we get the simplest non-trivial
topological space
%(수식1)
\begin{equation}
\label{eq1}
T_0 = \{\phi, F_{1,\bar{1}}, F^{2}_{1,\bar{1}},\cdots,F^{n}_{1,\bar{1}},\cdots\},
\end{equation}
where the subscript 0 is assigned to represent that there exist no
3-junctions in the set. However, if we include the gluonic flux-tube
$F_0$ as an excited component into $F_{1,\bar{1}}$, the counting
rule for 3-junctions has to be changed. Pairs of incoming and
outgoing 3-junctions can be added or removed indefinitely. Therefore
we will not consider this possibility in this paper. Then there
exists only one kind of flux-tube $F_{1,\bar{1}}$ in the space
$T_0$, so we may reduce the notation as
%(수식 2)
\begin{equation}
\label{eq2}
T_0 = \{\phi, F_{1,\bar{1}}\},
\end{equation}
where it is assumed that $F_{1,\bar{1}}$ can be multiplied
repeatedly without violating the law of baryon number conservation.
Now, the topological space for baryon-meson system
becomes
%(수식3)
\begin{equation}
\label{eq3}
T_1 = \{\phi, F_3\}
\end{equation}
with one incoming 3-junction multiplied by meson flux-tubes, which
is implicitly assumed. The next baryon-meson-baryon space can be
represented as
%(수식4)
\begin{equation}
\label{eq4}
T_2 = \{\phi, F^2_3\}.
\end{equation}
When outgoing 3-junctions exist, we need another index to represent
the topological space, for example, the space with two incoming
3-junctions and one outgoing 3-junction can be denoted as
%(수식 5)
\begin{equation}
\label{eq5} T_{2,\bar{1}} = \{\phi,
{F^2_3}F_{\bar{3}},{F_3}F_{2,\bar{2}},F_{4,\bar{1}} \}.
\end{equation}
In general, we can write down the spaces as
%(수식6)
\begin{equation}
\label{eq6}
T_{i,\bar{j}} = \{\phi, {F^i_3}F^j_{\bar{3}},{F^{i-1}_3}{F^{j-1}_{\bar{3}}}F_{2,\bar{2}},\cdots \},
\end{equation}
where $i$ is the number of incoming 3-junctions and $j$ is that of
outgoing 3-junctions. For a given space, the baryon number $B=i-j$
is fixed and the number of boundary points is reduced by (1,1) pair
by one union operation.\par
 For the constructed topological spaces, we can try to define
physical amplitude that is appropriate for the description of
gluonic degrees of freedom. The physical amplitude has to be defined
in such a way that it can be used to deduce quantitative predictions
about the behaviors of flux-tubes. Since the formation of a space is
generated by the union and intersection operations, the amplitude
can be taken to be related to the connection and disconnection of
flux-tubes. One method to define the amplitude is to consider the
connectedness between quarks through given flux-tube open set. Thus
let's introduce the amplitude $A$ for a quark to be connected to
another quark or antiquark. We can name it as connection amplitude.
If we can devise a formalism that provides quantitative descriptions
of $A$, we may use it to predict gluonic behaviors pictured as
flux-tubes.\par
 In order to devise quantitative descriptions of $A$, we need to consider
the relationships between $A$ and some physically measurable
quantity. Since $A$ is defined between two quarks, one possible
quantity can be chosen as the distance between the two quarks.
However, the measurement of distance between quarks cannot be
carried out without considering the gluonic interactions.
Measurements can be done either by scattering processes or by
analyzing bound state spectra. In either case, gluonic states affect
the measured value, and therefore, the definition of distance
between quarks has to be related in some way to gluonic properties.
Since we have introduced flux-tube open sets, it is plausible to
relate the measurement of distance to characteristics of open sets.
As we assigned connection amplitude $A$ to flux-tube open set, we
can consider general relationships between $A$ and measurement of
distance. These relationships can be formulated by assuming the
existence of a measure $M$ of $A$ satisfying the conditions
%(수식 7)
\begin{equation}
\label{eq7}
\begin{array}{ll}
1.& M(A)~{\rm{decreases~ as}}~ A~ \rm{increases},\\
2.& M(A_1)+M(A_2)=M({A_1}{A_2})~{\rm{when}}~ A_1~ {\rm{and}}~ A_2~
\rm{are~ independent}.
\end{array}
\end{equation}
The measurement of distance is generalized as measure $M$ and the
first condition states that the connection probability increases for
smaller measure of flux-tube. The other condition states the
relation between two flux-tubes that can be joined to form single
flux-tube or vice versa. With these two conditions, we can solve the
measure $M$ as functions of $A$
%(수식 8)
\begin{equation}
\label{eq8}
M(A)=-k \ln \frac{A}{A_0},
\end{equation}
where $A_0$ is a normalization constant and $k$ is appropriate
parameter. Now we can consider the measure $M$ as a metric function
defined between two quarks. For the simplest flux-tubes
$F_{1,\bar{1}}$, a metric function can be introduced between the two
boundary points $\vec{x}$ and $\vec{y}$ corresponding to the
positions of a quark and an antiquark. A general form of distance
function between the two points $\vec{x}$ and $\vec{y}$ can be
written down as $|\vec{x}-\vec{y}|^\nu$ with $\nu$ being an
arbitrary number. This distance function can be made metric for the
points $\vec{z}$ satisfying the condition
%(수식 9)
\begin{equation}
\label{eq9}
|\vec{x}-\vec{z}|^{\nu} + |\vec{z}-\vec{y}|^{\nu} \geqq |\vec{x}-\vec{y}|^{\nu}.
\end{equation}
For the points not satisfying this triangle inequality, we cannot
measure the distance from boundary points with given $\nu$. Then it
is possible to define the inner part of flux-tube as the set of
points contradicting the condition in Eq. (\ref{eq9}). With this
assignment, we can figure out the shape of flux-tube and we can take
$|\vec{x}-\vec{y}|^\nu$ as an appropriate measure to deduce a
concrete form for the connection amplitude $A$.\par
\begin{figure}[h]
\centering\includegraphics[width= 5cm, height=0.8cm ]{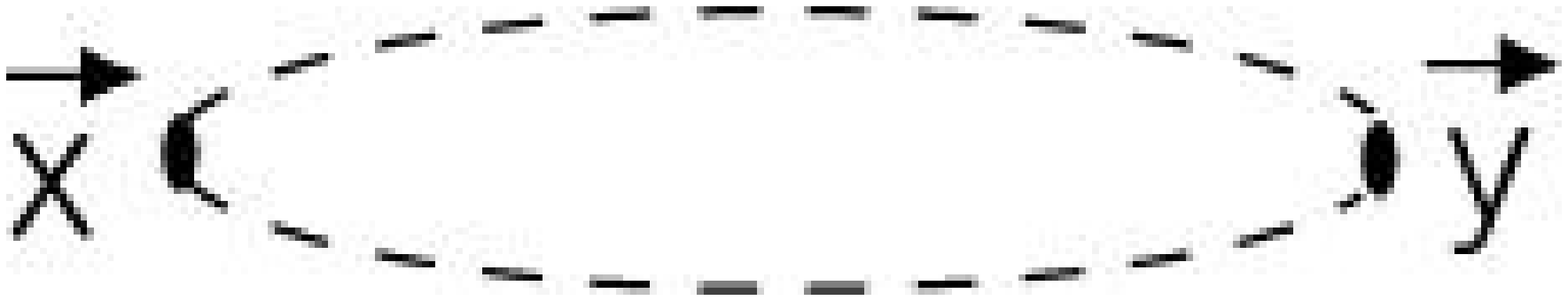}
\vspace{-2mm} \caption{\small{Flux-tube shape for $1<\nu<2$ with
boundaries at $\vec{x}$ and $\vec{y}$.} } \label{fig2}
\end{figure}

 In general, the value of $\nu$ is arbitrary, but the lower limit can
be fixed to be 1 because there exists no point $\vec{z}$ satisfying
the triangle inequality with $\nu<1$. For $\nu=2$ case, the shape of
flux-tube becomes sphere type, and the shape changes into concave
one if $\nu>2$. Since we consider the gluonic flux-tube as a smooth
structure, we can restrict the value of $\nu$ as
%(수식 10)
\begin{equation}
\label{eq10}
1 \leqq \nu \leqq 2.
\end{equation}
For a $\nu$ value between 1 and 2, we can draw the shape of
flux-tube as in Fig.\ref{fig2}. As the value of $\nu$ changes from 1
to 2, the flux-tube shape changes from a line into a sphere. In
order to account for various possibilities, we need to sum over
contributions from different ${\nu}'s$.
% 그림 2
For a small increment $d\nu$, the product of the two probability
amplitudes for $|\vec{x}-\vec{y}|^\nu$ and
$|\vec{x}-\vec{y}|^{{\nu}+d\nu}$ to satisfy the metric conditions
can be taken as the probability amplitude for the increased region
to be added into the inner connected region which is out of the
metric condition. Then, the full connection amplitude becomes
%(수식 11)
\begin{equation}
\label{eq11}
A = A_0 \exp \{{-\frac{1}{k}} \int^\alpha_1 F(\nu) r^\nu d\nu \} ,
\end{equation}
where all possibilities from the line shape with ${\nu}=1$ to the
arbitrary shape with ${\nu}=\alpha$ have been included. The weight
factor $F(\nu)$ is introduced to account for possible different
contributions from different ${\nu}'s$, and the variable $r$ is
%(수식 12)
\begin{equation}
\label{eq12}
r = \frac{1}{l} |\vec{x} - \vec{y}|
\end{equation}
with $l$ being a scale parameter. If we sum up to the spherical
shape flux-tube with ${\alpha}=2$, we get in case of equal weight
$F(\nu)=1$
%(수식 13)
\begin{equation}
\label{eq13}
A=A_0 \exp \{{-\frac{1}{k}} \frac{r^2 - r}{\ln r}\}.
\end{equation}
In fact, we can generalize this form by replacing
%(수식 14)
\begin{equation}
\label{eq14}
A(r)=r^a B(r)
\end{equation}
in Eq. (\ref{eq7}) which is satisfied with $a>0$. Then the new form
of connection amplitude becomes
%(수식 15)
\begin{equation}
\label{eq15} B(r)=\frac{B_0}{r^a} \exp \{{-\frac{1}{k}} \frac{r^2 -
r}{\ln r}\},
\end{equation}
and we will use this form in this paper to describe the gluonic
structures of tetraquarks.
%3절
%
\section{Calculations of Gluonic Structures for Tetraquarks}
\label{three}
 The flux-tube overlap function $\gamma$ represents the probability amplitude
for flux-tubes to be overlapped before and after the change of
flux-tube configurations\cite{R3}. It is introduced to predict the
position of new quark pair which generates the changes of flux-tube
boundaries. These predictions are essential in estimating the
amplitudes for strong decays of hadrons. In our formalism, the
function $\gamma$ is given by the product of the initial and the
final connection amplitudes
%(수식 16)
\begin{equation}
\label{eq16}
\gamma = A_i A_f ,
\end{equation}
where $A_i$ and $A_f$ represent the connection amplitudes before and
after quark pair creation. Since the quark pairs are created by the
gluons in hadrons, the gluonic contents of hadrons can be probed by
the probability amplitudes for quark pair creations, which are
described by the flux-tube overlap function $\gamma$. The
equi-$\gamma$ curves can be taken as representing the gluonic
structures of hadrons or other multiquark states.\par
\begin{figure}[h]
\centering\includegraphics[width=50mm,height=30mm]{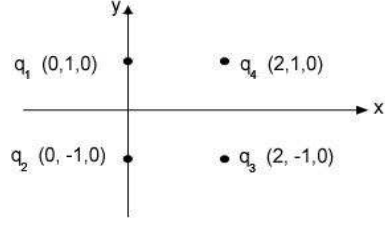}
\vspace{-2mm} \caption{\small{Four positions of quarks.} }
\label{fig3}
\end{figure}
\begin{figure}[h]
 \centering \subfigure[]
  {\includegraphics[width=40mm,height=30mm]{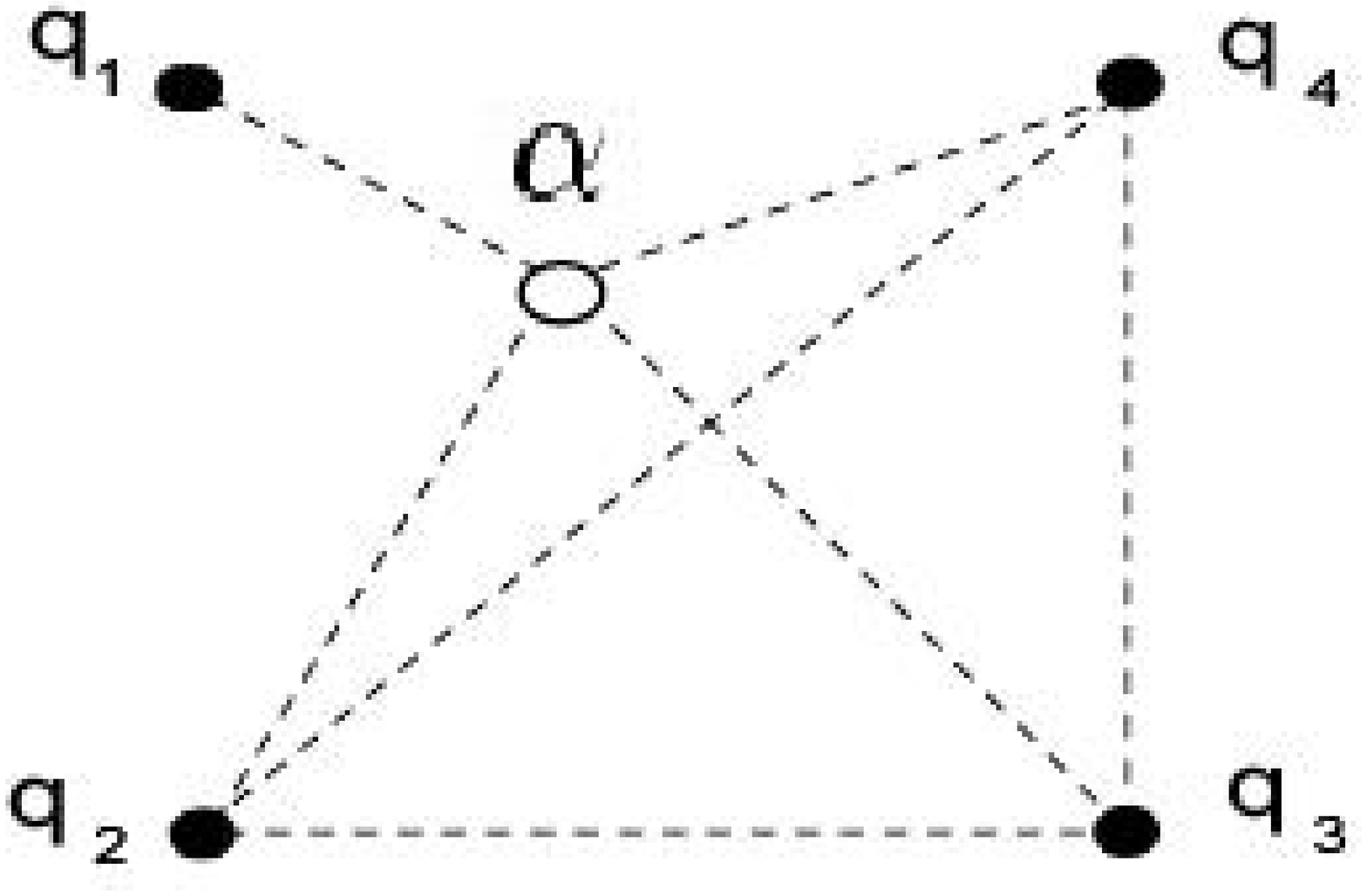}}
  \hspace{10mm}
 \centering\subfigure[]
  {\includegraphics[width=40mm,height=30mm]{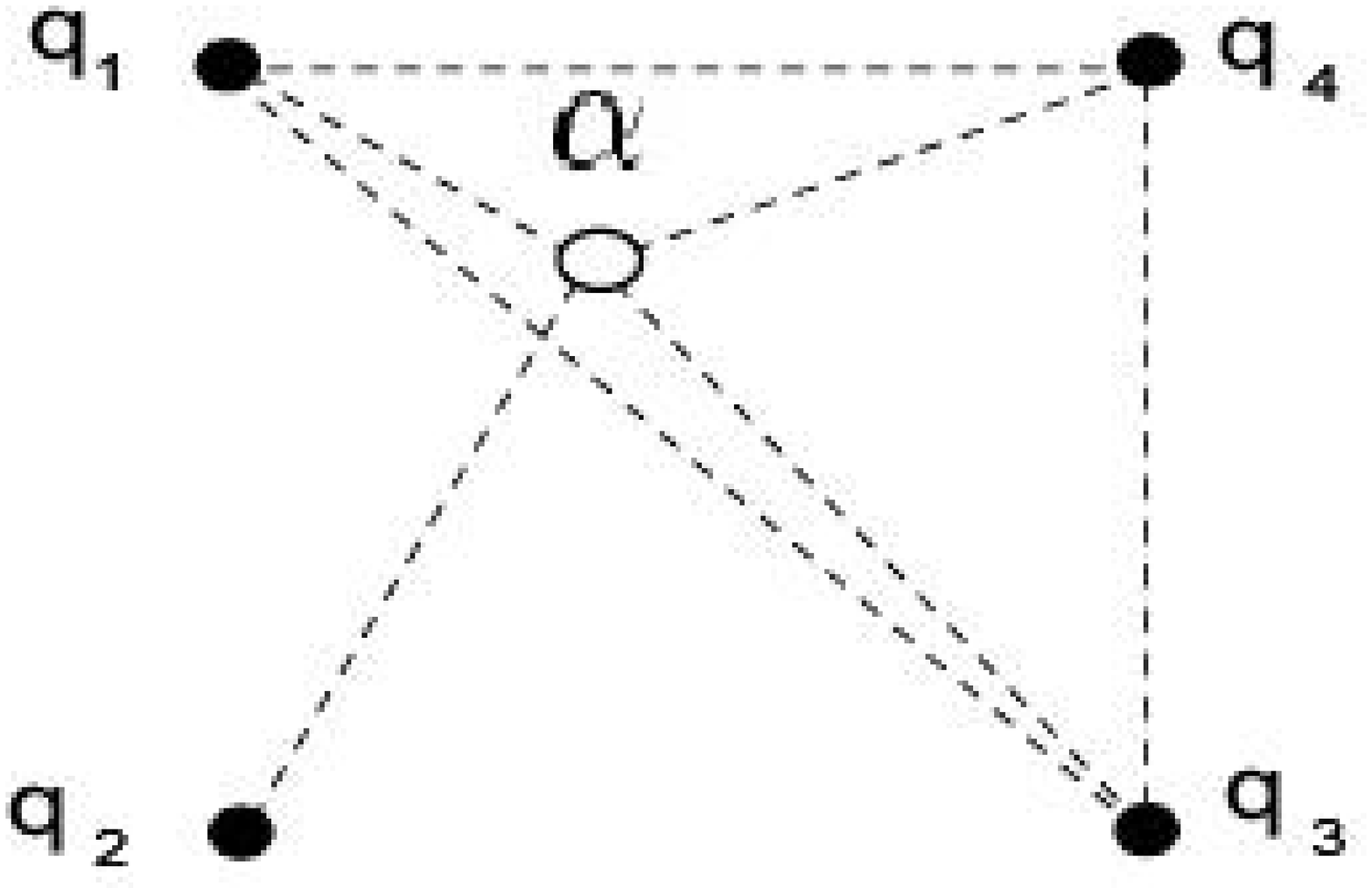}}
  \hspace{10mm}
    \centering\subfigure[]
  {\includegraphics[width=40mm,height=30mm]{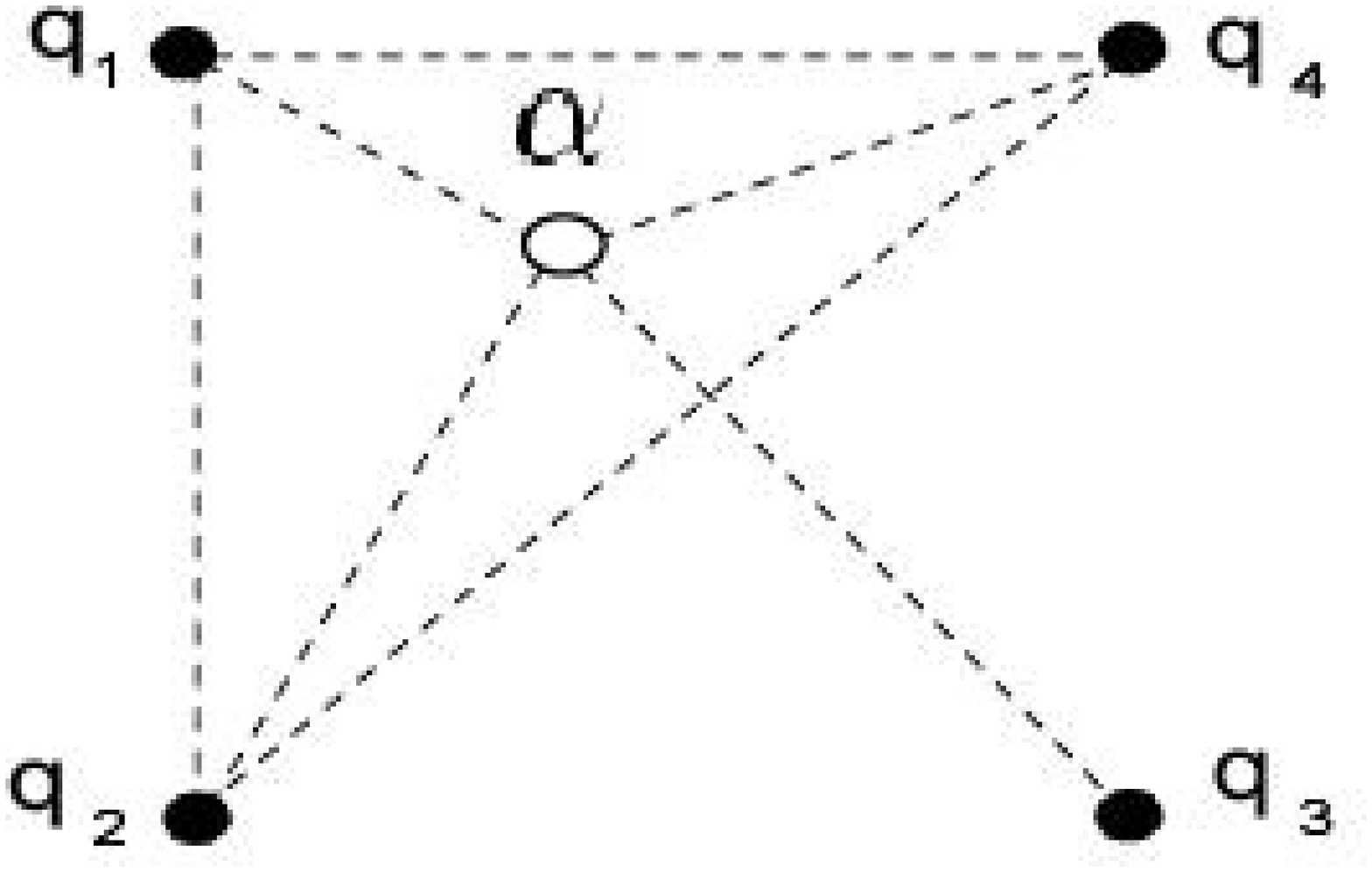}}
  \vspace{5mm}
  \centering\subfigure[]
  {\includegraphics[width=40mm,height=30mm]{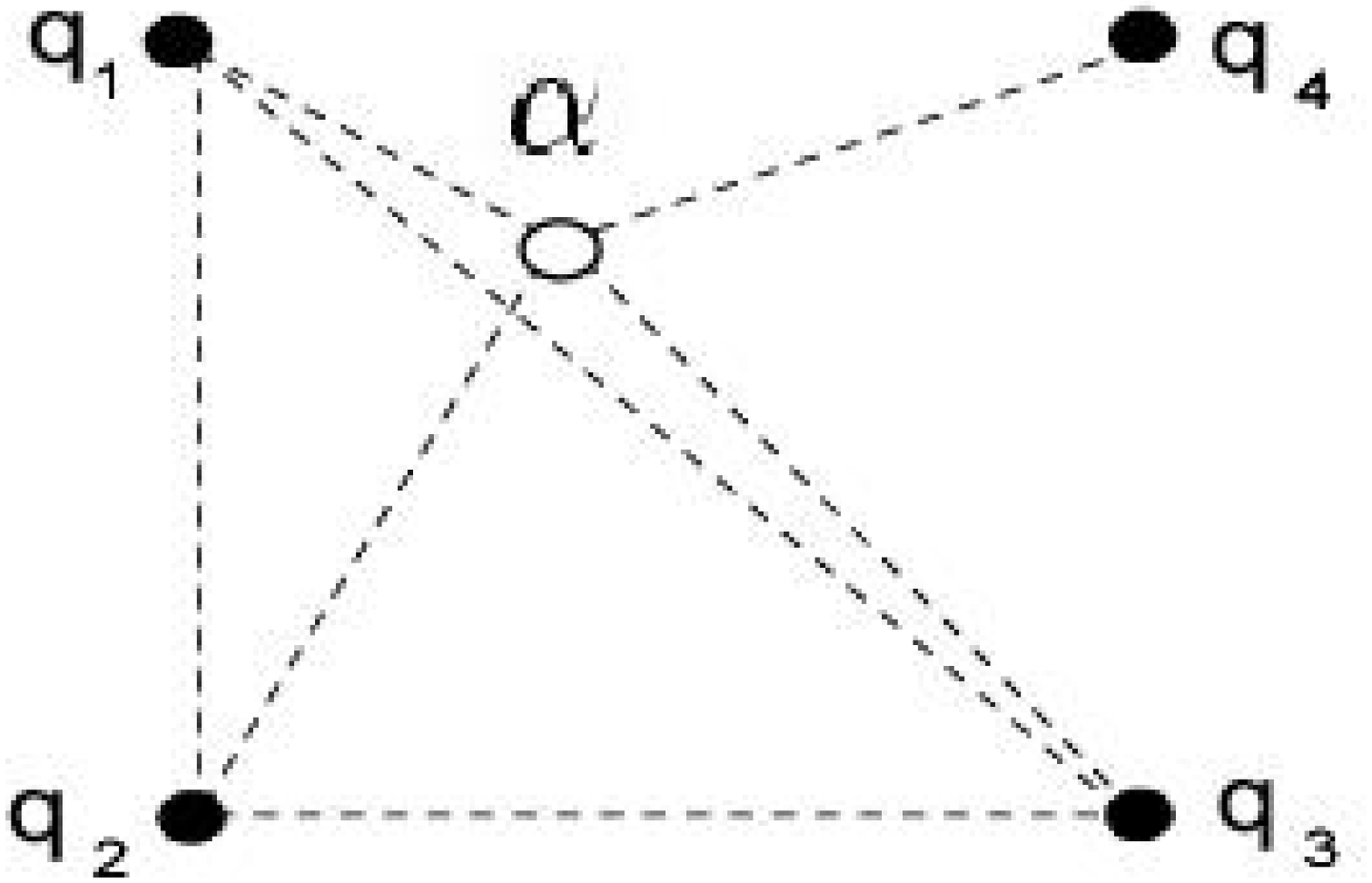}}
   \hspace{22mm}
     \subfigure[]
  {\includegraphics[width=40mm,height=30mm]{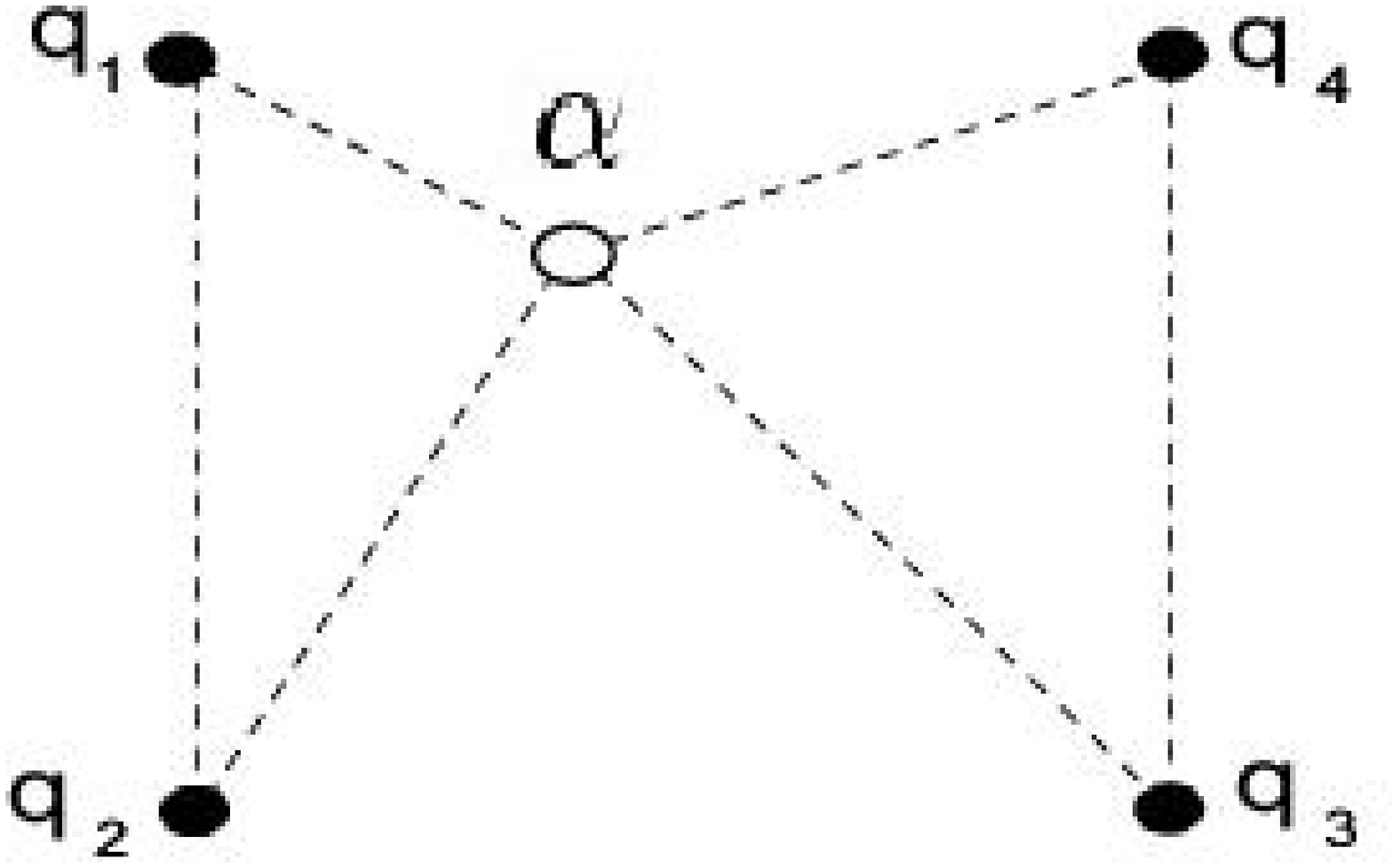}}
   \caption{\small{Possible new flux-tube connections with quark pair created at
 $\alpha$(white circle).}}
 \label{fig4}
\end{figure}

 Now let's calculate the equi-$\gamma$ curves for tetraquarks. Since the
relative positions of four quarks are unknown, we consider the two
cases of rectangular shape and tetrahedron shape. Rectangular
configuration is typical of plane structure and tetrahedron is the
simplest one in 3-dimension. For planar configuration, let's take
the quark positions at (0,1,0), (0,-1,0), (2,-1,0), and (2,1,0) as
shown in Fig.\ref{fig3}.
% 그림 3
Then the initial connection amplitude  becomes
%(수식 17)
\begin{equation}
\label{eq17}
A_i = A_{12} A_{13} A_{14} A_{23} A_{24} A_{34} .
\end{equation}
Since this amplitude is fixed once the quark positions are given,
the overlap function $\gamma$ is not affected by the value of $A_i$.
However, in case of moving quarks, $A_{i}$ will contribute to the
final form of $\gamma$. In our case, $A_{i}$ is just a factor of
normalization. Instead the final connection amplitude $A_{f}$
depends on the position of created quark pair, and the possible
flux-tube connections are shown in Fig.\ref{fig4}.\par
% \vspace{5mm}
% 그림4
The four cases from Fig.\ref{fig4}(a) to Fig.\ref{fig4}(d)
correspond to one meson and one tetraquark, whereas the case of
Fig.\ref{fig4}(e) corresponds to one baryon and one antibaryon. For
Fig.\ref{fig4}(a), we can write down the $A_{f}$ as
%(수식 18)
\begin{equation}
\label{eq18}
A^a_f = A^a_0 \prod^4_{i=1} \frac{1}{r^a_{{\alpha}{i}}}
       \exp \Big\{ {-\frac{1}{k}} \frac{r^2_{{\alpha}{i}} - r_{{\alpha}{i}}} {\ln r_{{\alpha}{i}}} \Big\}
       \times
       \prod_{i,j\neq 1} \frac{1}{r^a_{{i}{j}}}
       \exp \Big\{ {-\frac{1}{k}} \frac{r^2_{{i}{j}} - r_{{i}{j}}} {\ln r_{{i}{j}}} \Big\}
\end{equation}
by using the form of connection amplitude given in Eq.(\ref{eq15}).
We can see easily that the amplitudes for Fig.\ref{fig4}(b), (c),
(d) cases are the same as in the form in Eq.(\ref{eq18}). On the
other hand, the amplitude for Fig.\ref{fig4}(e) case becomes
%(수식 19)
\begin{equation}
\label{eq19}
  A^e_f = A^e_0 \prod^4_{i=1} \frac{1}{r^a_{\alpha i}}
       \exp \Big\{ {-\frac{1}{k}} \frac{r^2_{\alpha i} - r_{\alpha i}} {\ln r_{\alpha i}} \Big\}
       \times
       \frac{1}{r^a_{12}}
       \exp \Big\{ {-\frac{1}{k}} \frac{r^2_{12} - r_{12}} {\ln r_{12}} \Big\}
       \times
       \frac{1}{r^a_{34}}
       \exp \Big\{ {-\frac{1}{k}} \frac{r^2_{34} - r_{34}} {\ln r_{34}} \Big\}
\end{equation}
with different normalization factor $A^e_0$. For fixed quark
positions, only the first factors in Eq.(\ref{eq18}) and
Eq.(\ref{eq19}) will contribute to the $\gamma$ structures, and with
appropriate replacements of normalization factors we get
%(수식 20)
\begin{equation}
\label{eq20}
 \gamma = \gamma_0 \prod^4_{i=1} \frac{1}{r^a_{\alpha i}}
          \exp \Big\{ -\frac{1}{k} \frac{r^2_{\alpha i} - r_{\alpha i}}{\ln r_{\alpha i}}
          \Big\}.
\end{equation}
There remain two parameters $k$ and $a$ and we we will fix the value
of $k$ as $k=1$ and calculate the $\gamma$ values for several values
of $a$.\par
\begin{figure}[h]
 \centering \subfigure[~$a=0.5$ case]
  {\includegraphics[width=49mm,height=45mm]{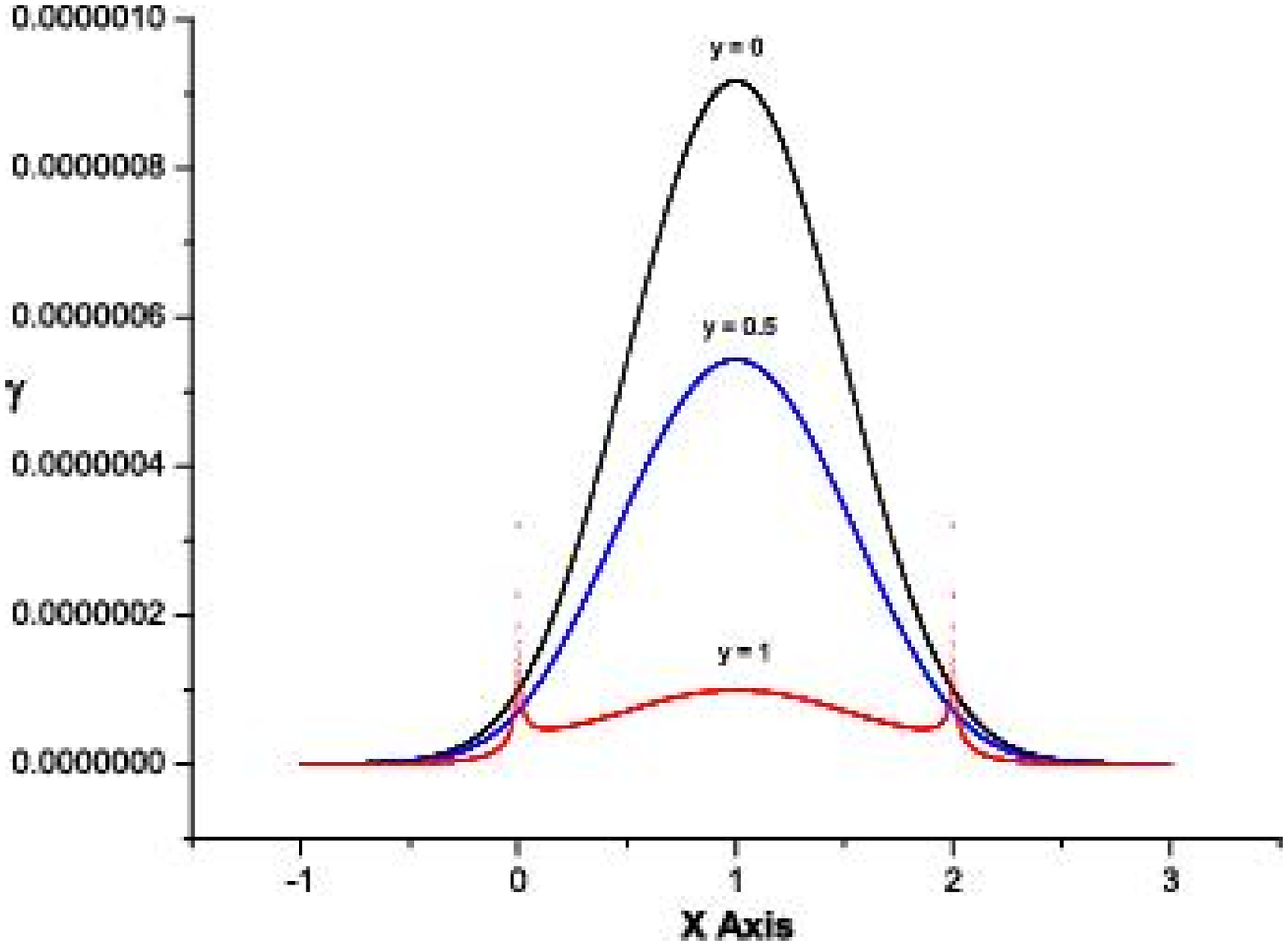}}
\centering \subfigure[~$a=1.0$ case]
 {\includegraphics[width=49mm,height=45mm]{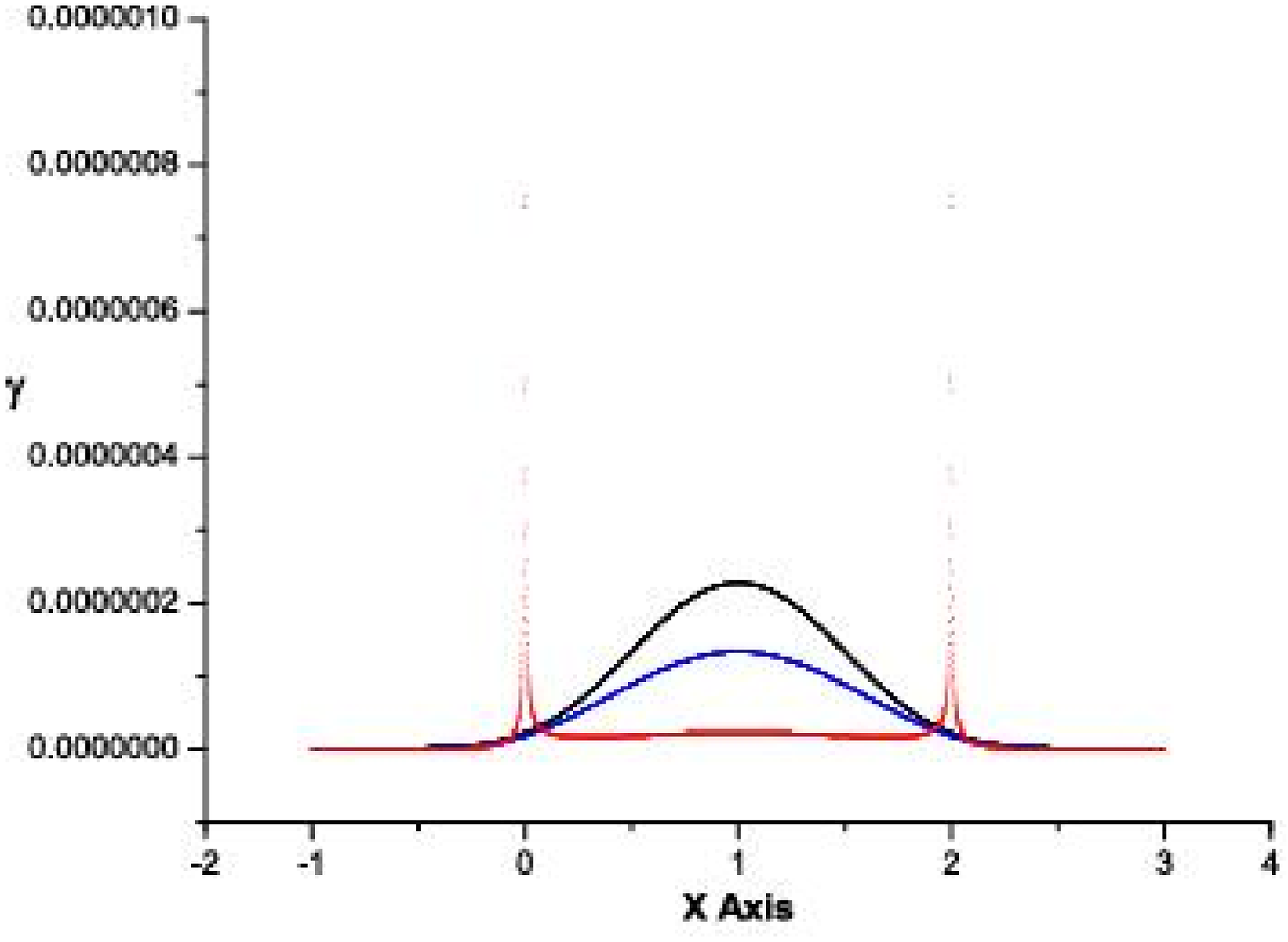}}
\subfigure[~$a=2$ case]
 {\includegraphics[width=49mm,height=45mm]{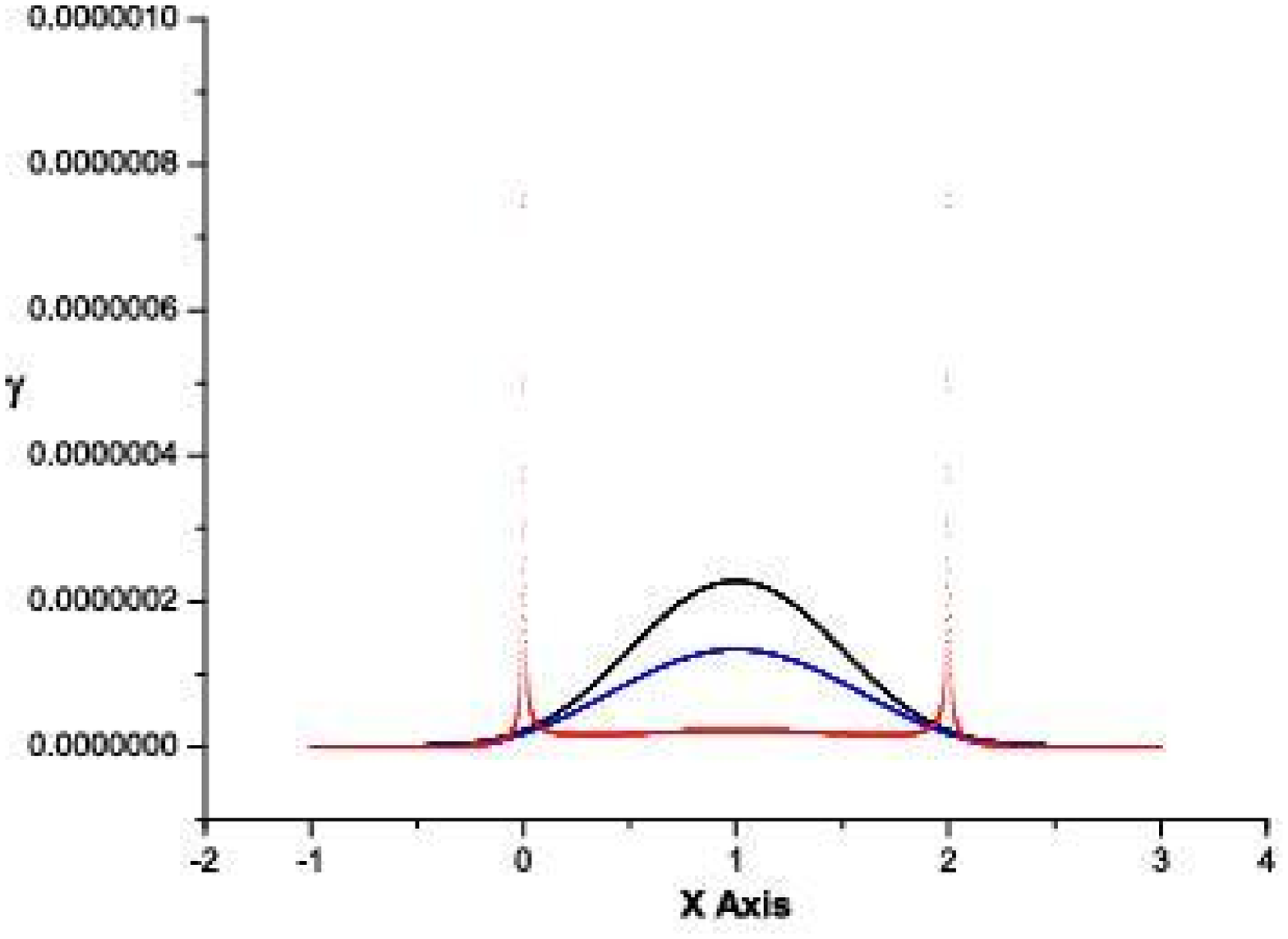}}
 \vspace{-2mm}
\caption{\small{Calculated overlap function $\gamma$ along the axes
parallel to $x$-axis.}} \label{fig5}
\end{figure}
\begin{figure}[h]
\centering \subfigure[~$a=0.5$ case]
 {\includegraphics[width=49mm,height=45mm]{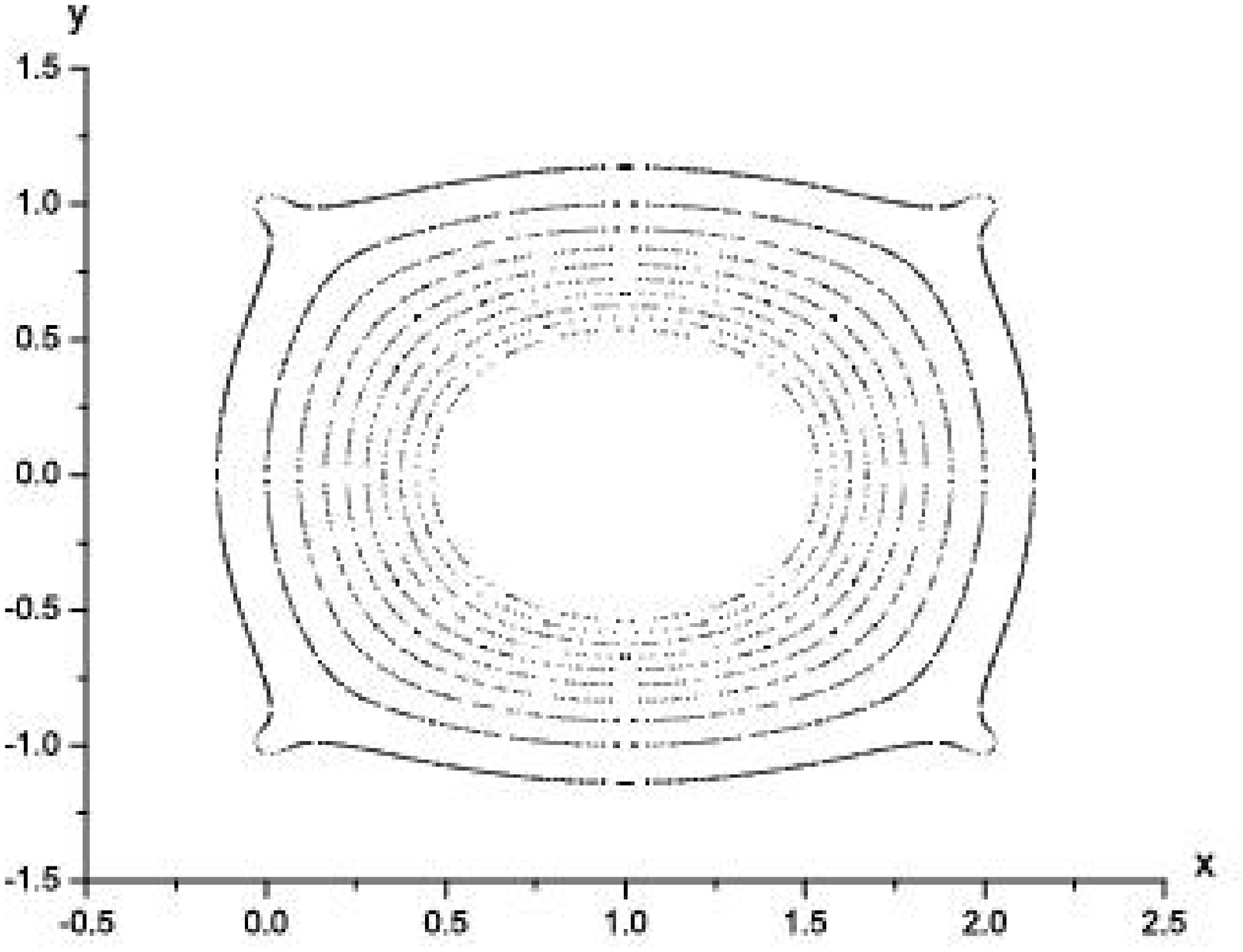}}
\centering\subfigure[~$a=1.0$ case]
 {\includegraphics[width=49mm,height=45mm]{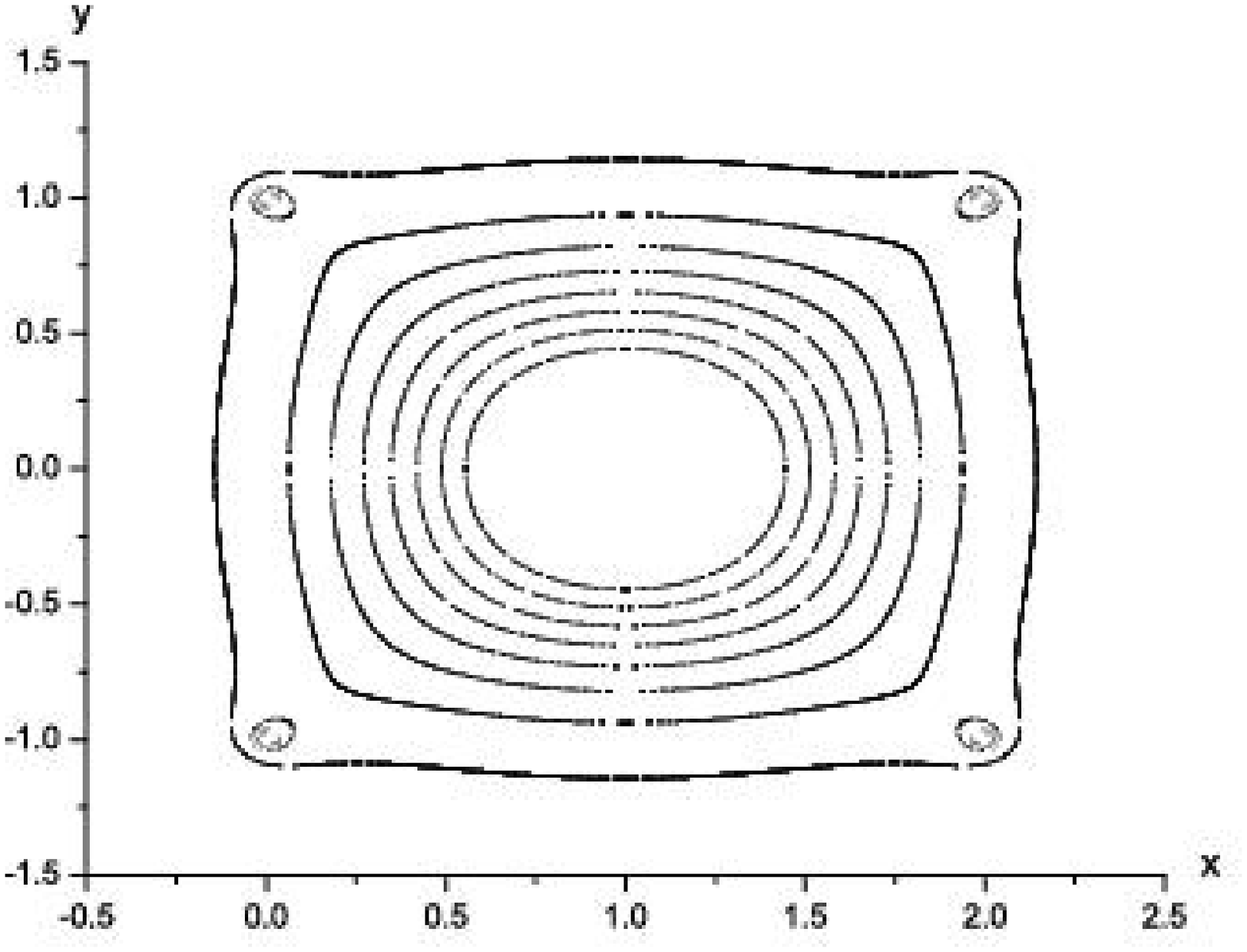}}
\subfigure[~$a=2$ case]
 {\includegraphics[width=49mm,height=45mm]{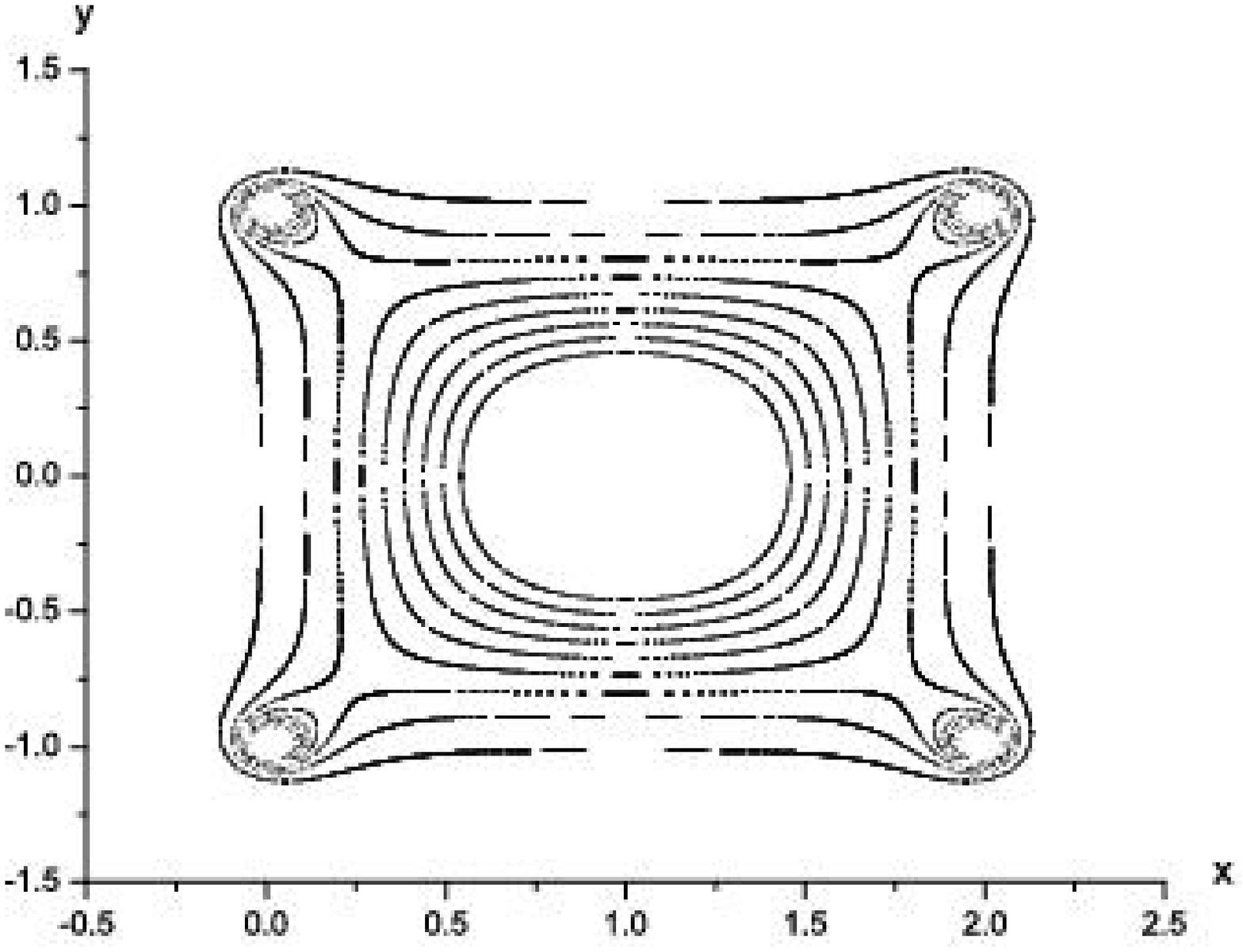}}
 \vspace{-2mm}
\caption{\small{Equi-$\gamma$ curves for different $a$ values.}}
\label{fig6}
\end{figure}
\begin{figure}[]
\centering \subfigure[~Outer equi-$\gamma$ curves]
 {\includegraphics[width=50mm,height=38mm]{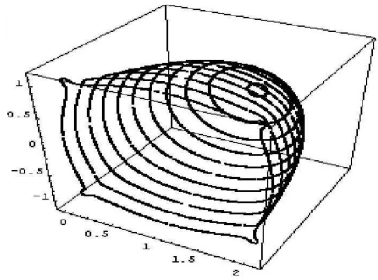}}
  \hspace{30mm}
\centering\subfigure[~Inner equi-$\gamma$ curves]
 {\includegraphics[width=50mm,height=38mm]{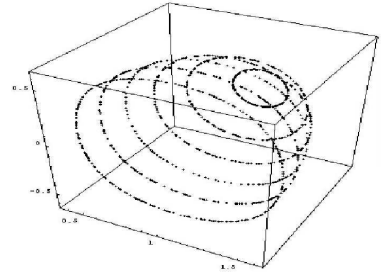}}
  \vspace{-2mm}
 \caption{\small{3-dimensional structures for $a=0.5$} }
\label{fig7}
\end{figure}
\begin{figure}[]
\centering \subfigure[~Outer equi-$\gamma$ curves]
 {\includegraphics[width=50mm,height=38mm]{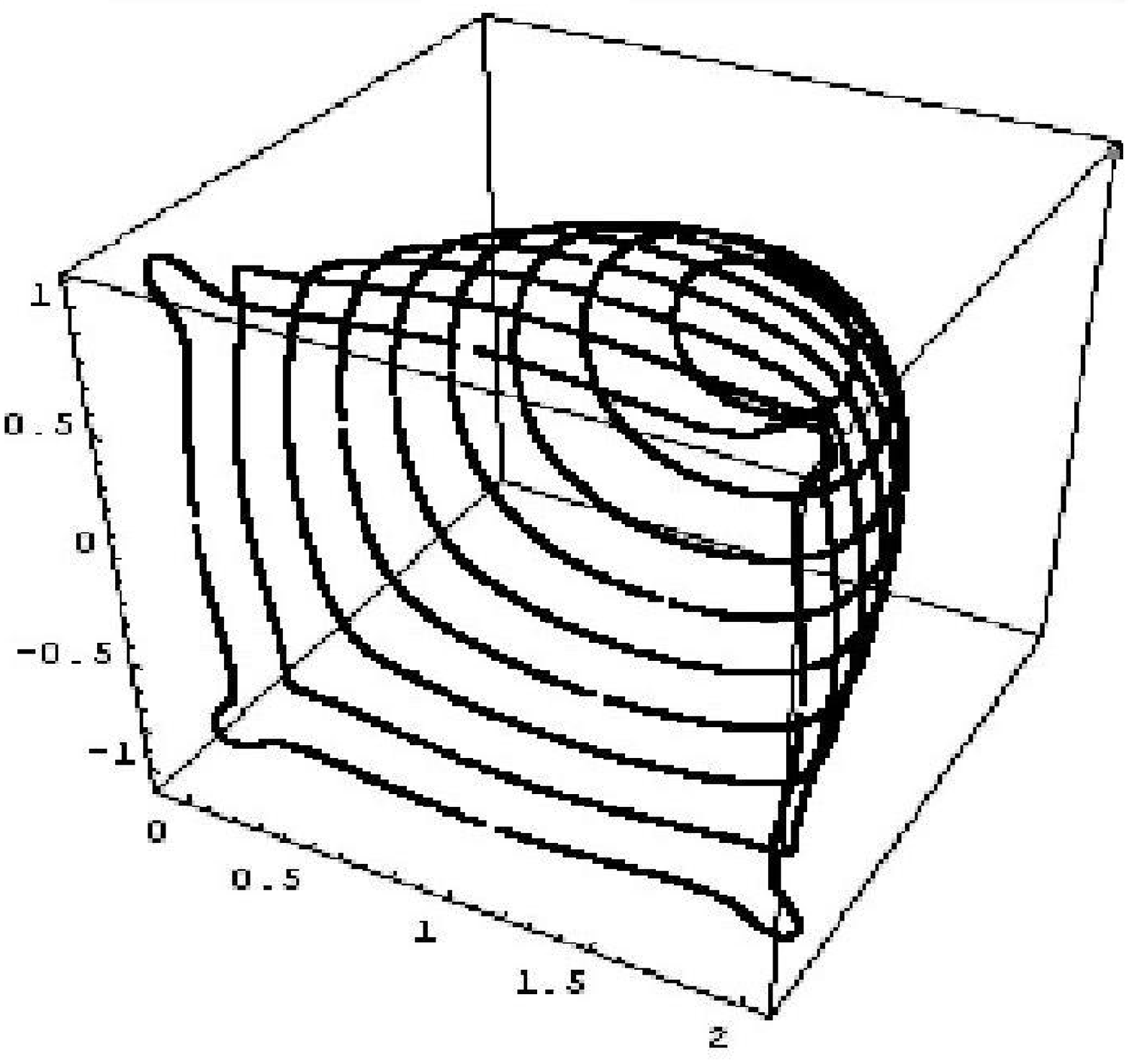}}
  \hspace{30mm}
\centering\subfigure[~Inner equi-$\gamma$ curves]
 {\includegraphics[width=50mm,height=38mm]{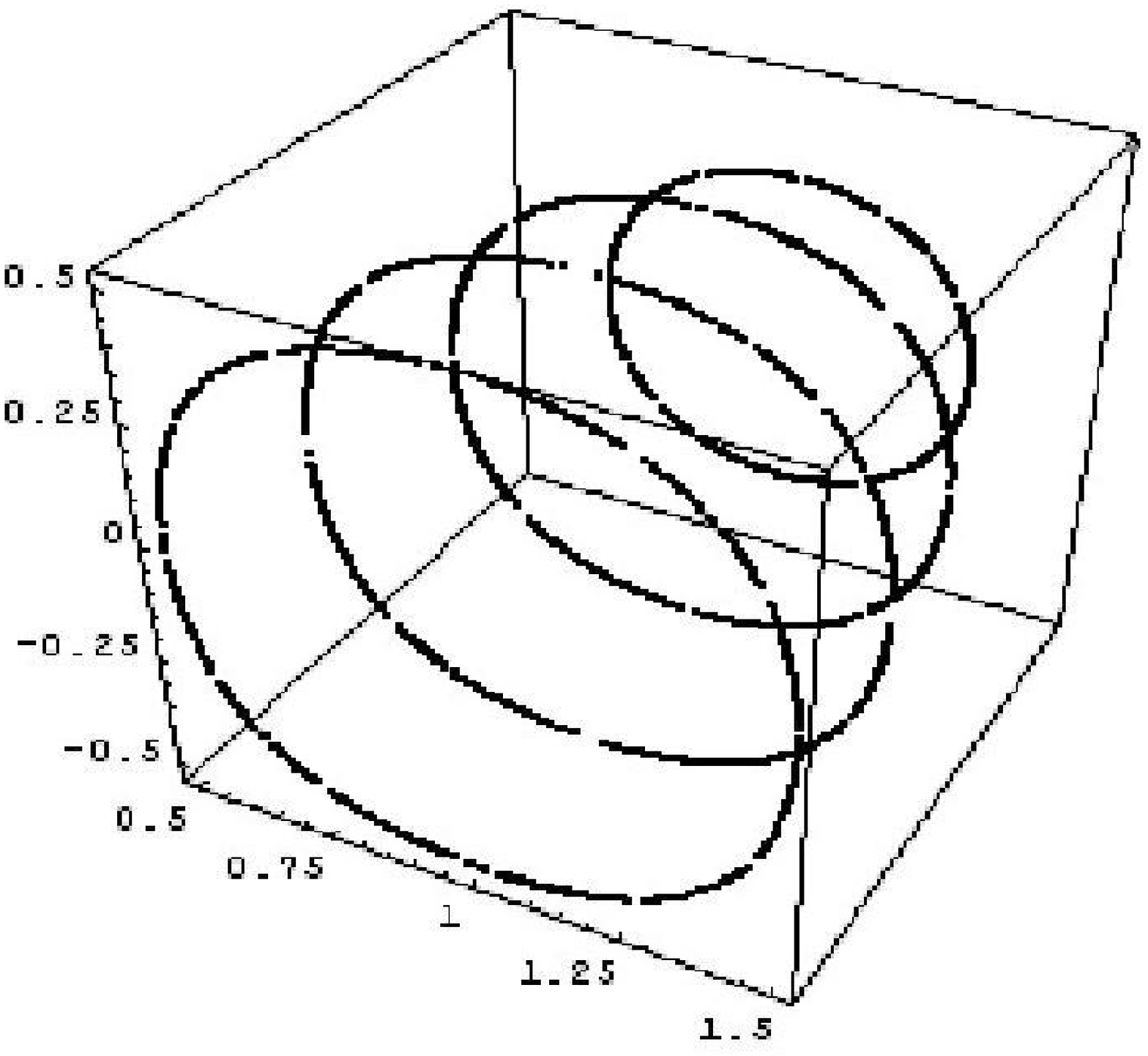}}
  \vspace{-2mm}
 \caption{\small{3-dimensional structures for $a=1.0$}}
 \label{fig8}
 \end{figure}
\begin{figure}[]
\centering \subfigure[~Outer equi-$\gamma$ curves]
 {\includegraphics[width=50mm,height=38mm]{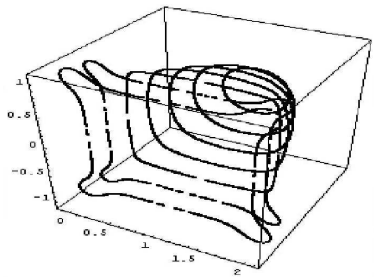}}
  \hspace{30mm}
\centering\subfigure[~Inner equi-$\gamma$ curves]
 {\includegraphics[width=50mm,height=38mm]{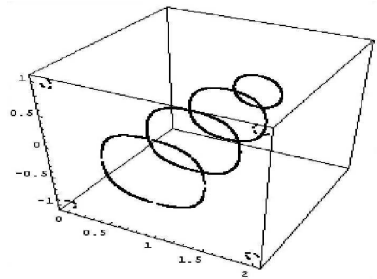}}
  \vspace{-2mm}
 \caption{\small{3-dimensional structures for $a=2.0$}}
 \label{fig9}
 \end{figure}

 In Fig.\ref{fig5}, we have presented the variations of $\gamma$ values along the $x$-axis.
We have taken the 3 cases of $y=0, 0.5, 1$ for each of the $a$
values $0.5, 1.0$, and $2.0$. The normalization factor $\gamma_{0}$
is set to arbitrary value to get comparable graphs for different
parameters. We can easily see that there appear high peaks at quark
positions with non-vanishing $a$ values. In Fig.\ref{fig6},
%그림 5
%그림 6
equi-$\gamma$ curves are shown for three $a$ values $a=0.5, 1.0$,
and $2.0$. We can find that gluonic densities are higher in central
region and just around the quarks. The singular behaviors at
boundary quarks can be removed by introducing some cutoff radius
$r_{cut}$, beyond which we can apply perturbative calculations
factorized from nonperturbative treatment of flux-tubes. We can
further draw 3-dimensional structures as in Fig.\ref{fig7},
Fig.\ref{fig8}, and Fig.\ref{fig9}.
%그림 7

%그림 8

%그림 9
\begin{figure}[]
\centering\includegraphics[width=60mm,height=50mm]{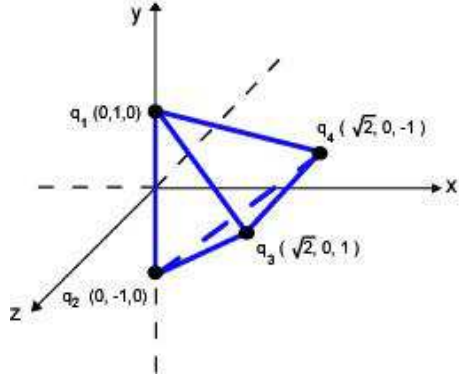}
\caption{\small{Four positions of quarks in tetrahedron. }}
\label{fig10}
\end{figure}

\begin{figure}[]
\centering \subfigure[~$a=0.5$ case]
 {\includegraphics[width=49mm,height=45mm]{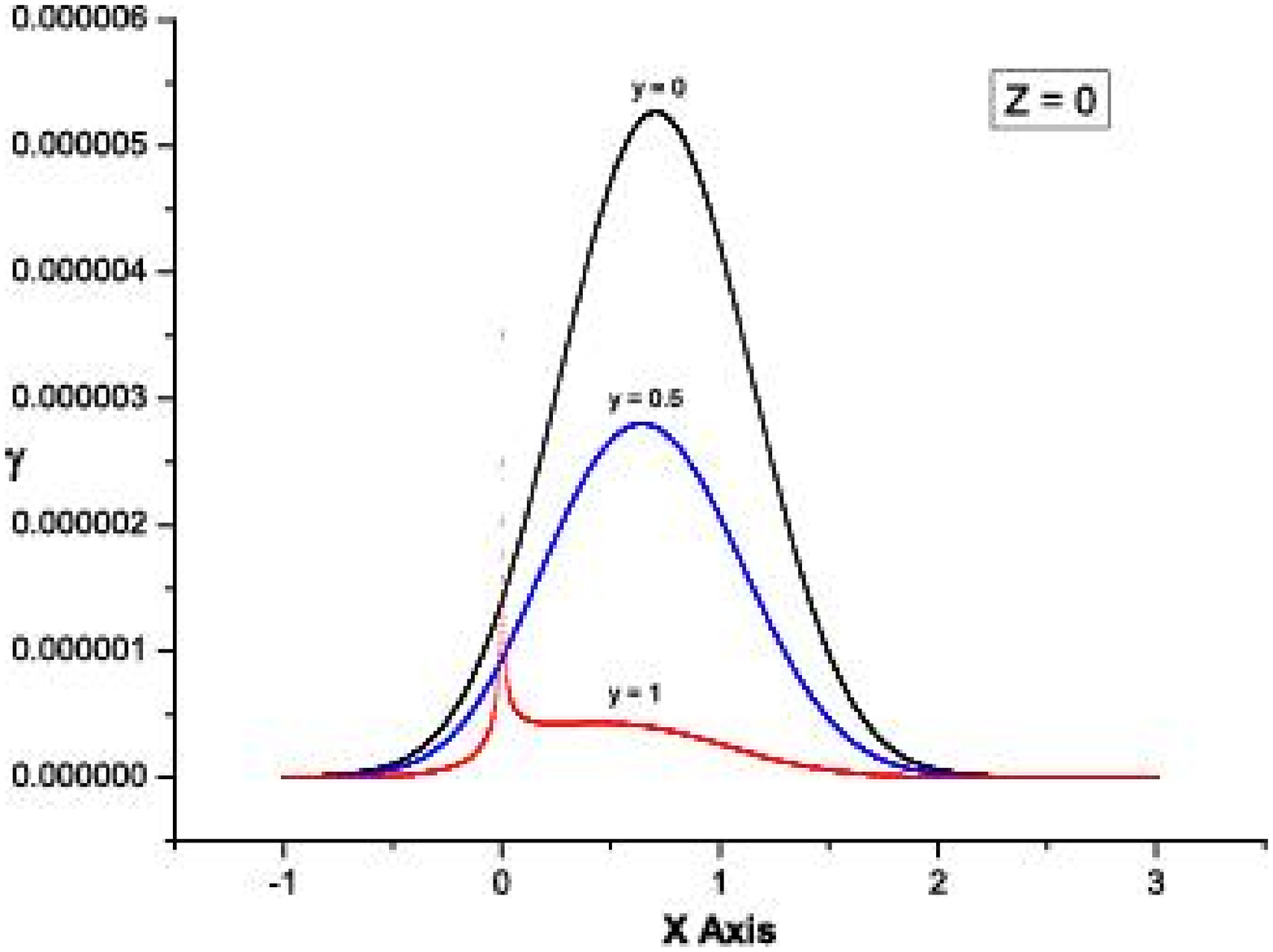}}
\centering\subfigure[~$a=1.0$ case]
 {\includegraphics[width=49mm,height=45mm]{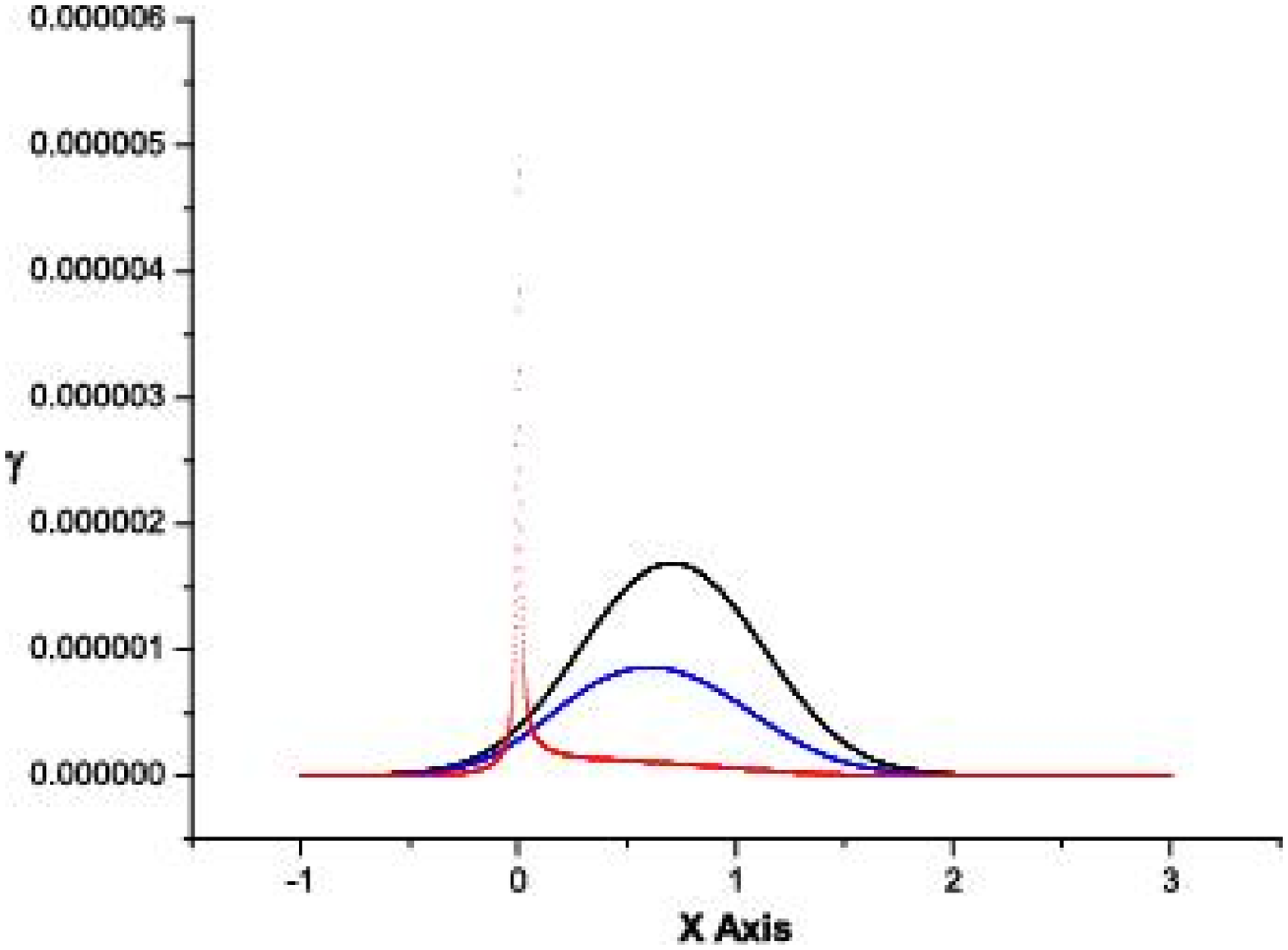}}
\subfigure[~$a=2$ case]
 {\includegraphics[width=49mm,height=45mm]{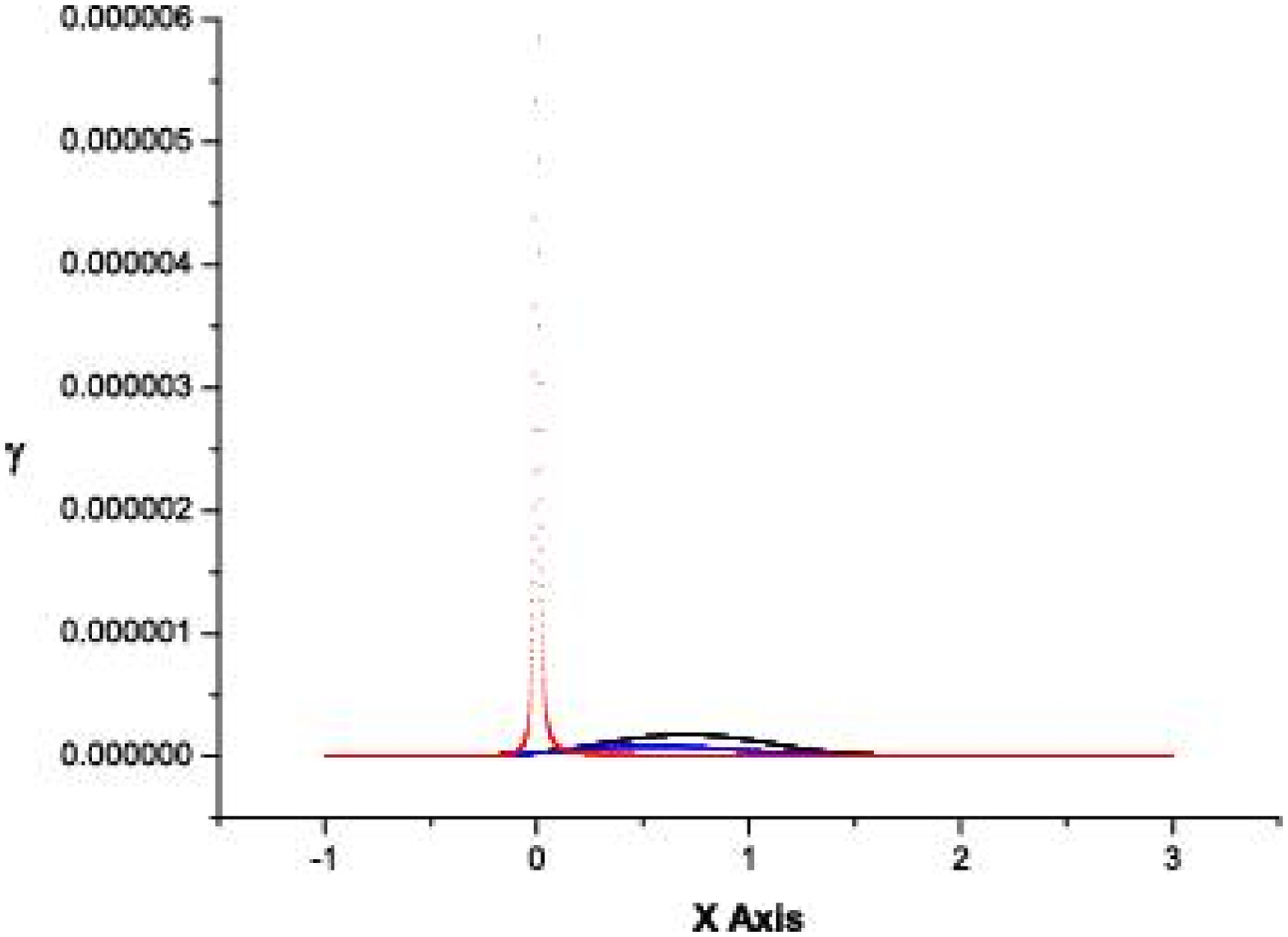}}
 \vspace{-2mm}
\caption{\small{Flux-tube overlap function $\gamma$ for tetrahedron
shape.}} \label{fig11}
\end{figure}

 Now let's turn to the case of tetrahedron shape. If we take the
four quark positions at (0,1,0),~(0,-1,0),~($\sqrt{2}$,0,1), and
 ($\sqrt{2}$,0,-1), the situation can be drawn as in Fig.\ref{fig10}.
 \vspace{6cm}
 % 그림 10
 The flux-tube overlap function $\gamma$ is of the same form as given
in Eq.(\ref{eq20}). With $k=1$, we can calculate $\gamma$ values for
different choices of $a$. The results are shown in Fig.\ref{fig11},
where $a$ values are taken to be 0.5, 1.0, and 2.0 respectively. The
three curves for each of $a$ correspond to the $y$ values 0, 0.5,
and 1, and the $z$ value is taken to be 0, i.e., at $xy$ plane. At
$x=1$, the largest $\gamma$ value appears at $y=0$, and then it
decreases as the value of $y$ increases. These features are the same
for most of $x$ values, however, there exist exceptions around $x=0$
and $y=1$ where a boundary quark is set to be fixed. We can easily
see a sharp peak around a boundary point, which can be taken as
implying that gluons gather just around a quark.
%그림 11
%(그림 12)
\begin{figure}[]
\centering \subfigure[~$a=0.5$ case]
 {\includegraphics[width=49mm,height=45mm]{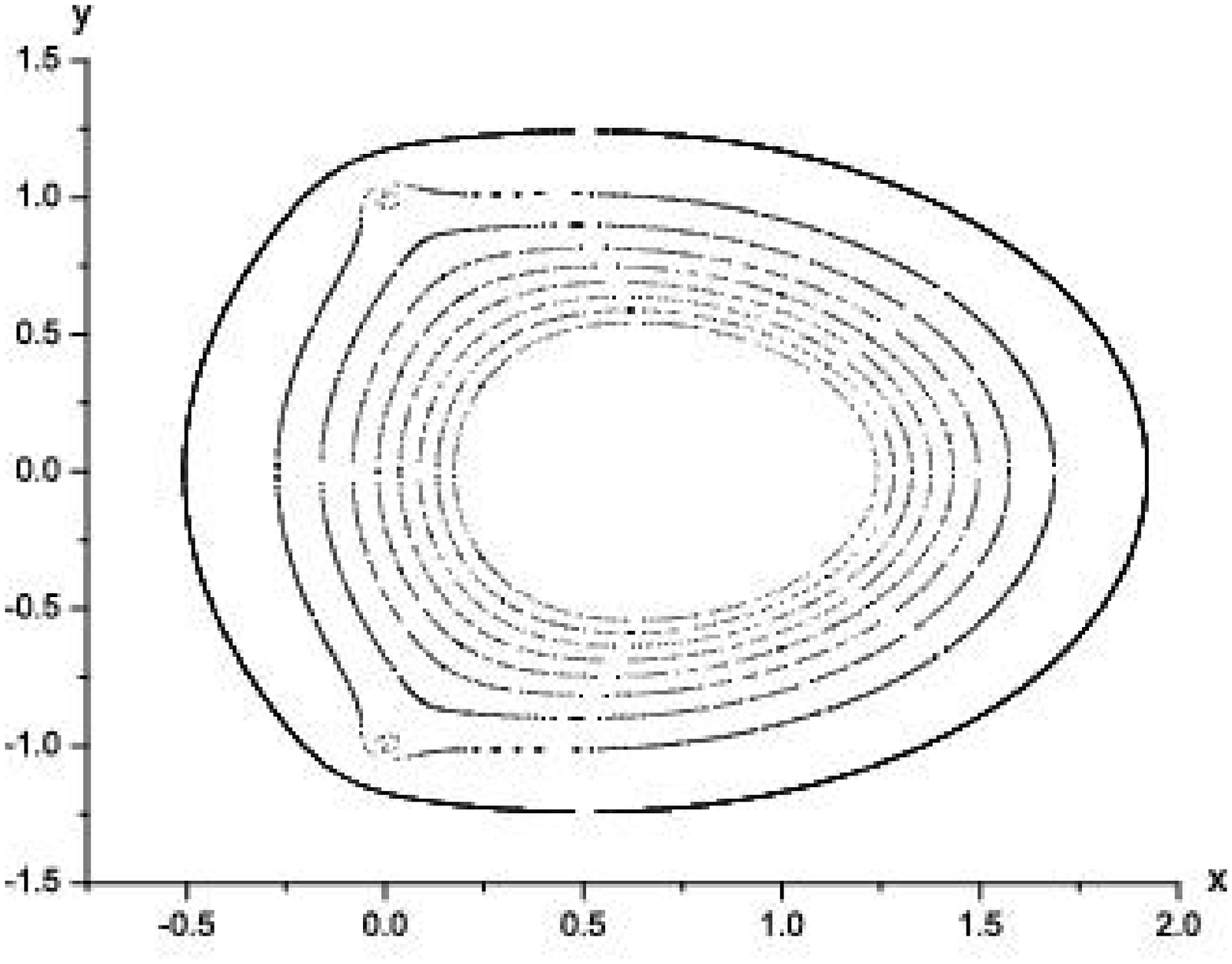}}
\centering\subfigure[~$a=1.0$ case]
 {\includegraphics[width=49mm,height=45mm]{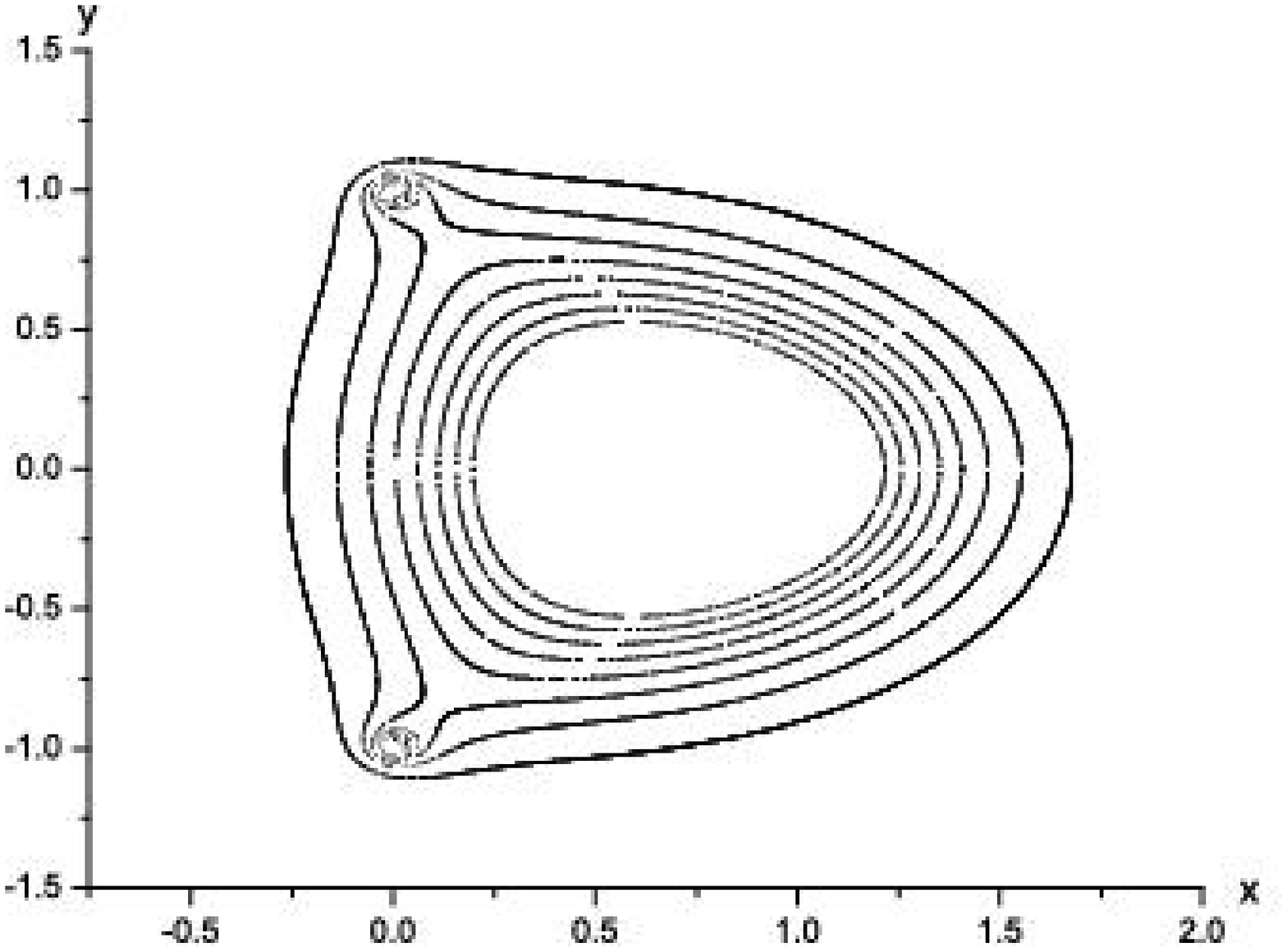}}
\subfigure[~$a=2$ case]
 {\includegraphics[width=49mm,height=45mm]{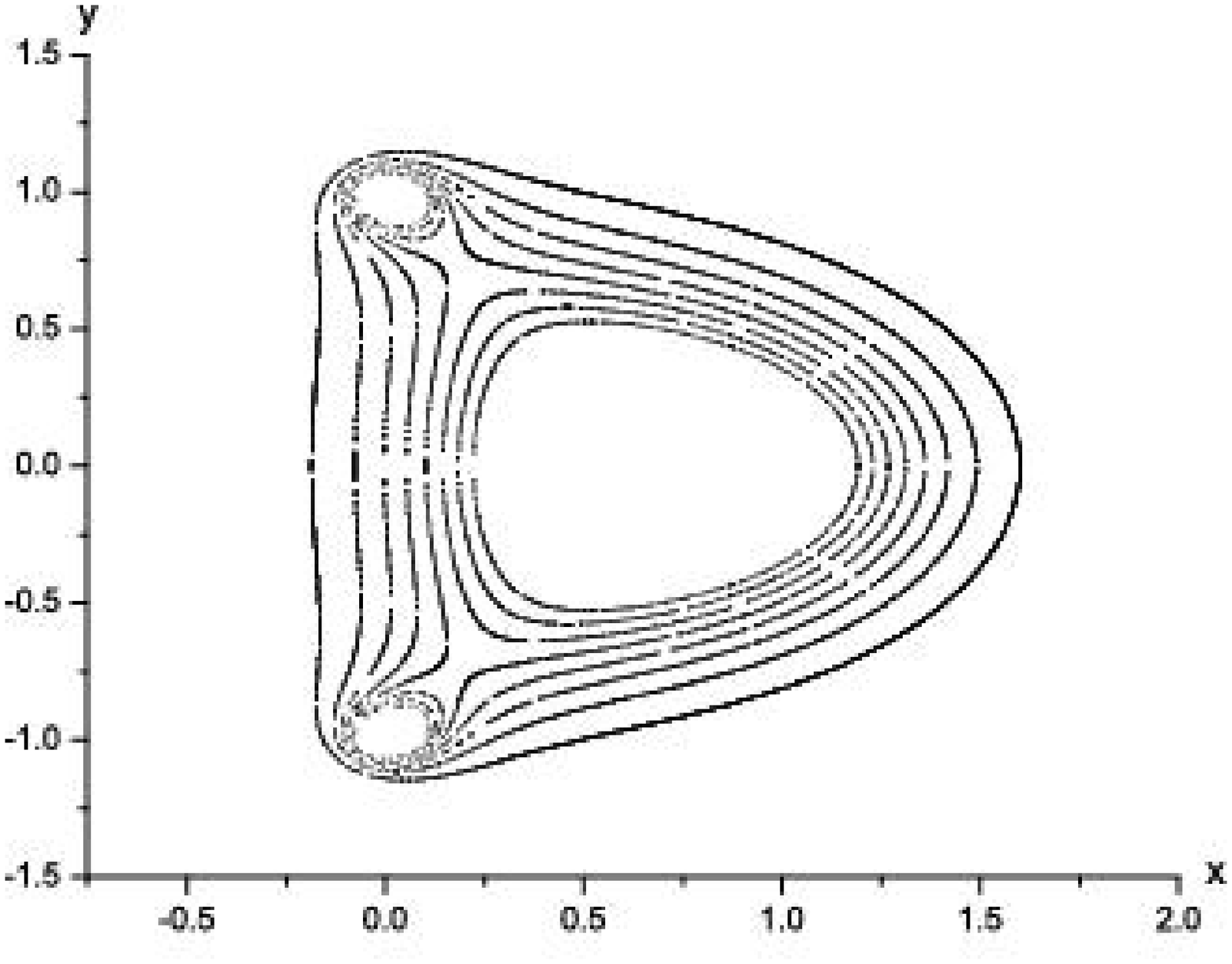}}
 \vspace{-2mm}
\caption{\small{Equi-$\gamma$ curves in $xy$ plane for different $a$
values.}} \label{fig12}
\end{figure}
For general structures, we  can draw the equi-$\gamma$ curves as in
Fig.\ref{fig12}. Again we can confirm the gathering of gluons around
each quark, and there appears a broad peak in central part
surrounded by four quarks. In order to figure out the situation in 3
dimension, we draw the equi-$\gamma$ curves in Fig.\ref{fig13},
Fig.\ref{fig14}, and Fig.\ref{fig15} with respective $a$ values.

% (그림13)
\begin{figure}[]
\centering \subfigure[~Outer equi-$\gamma$ curves]
 {\includegraphics[width=50mm,height=40mm]{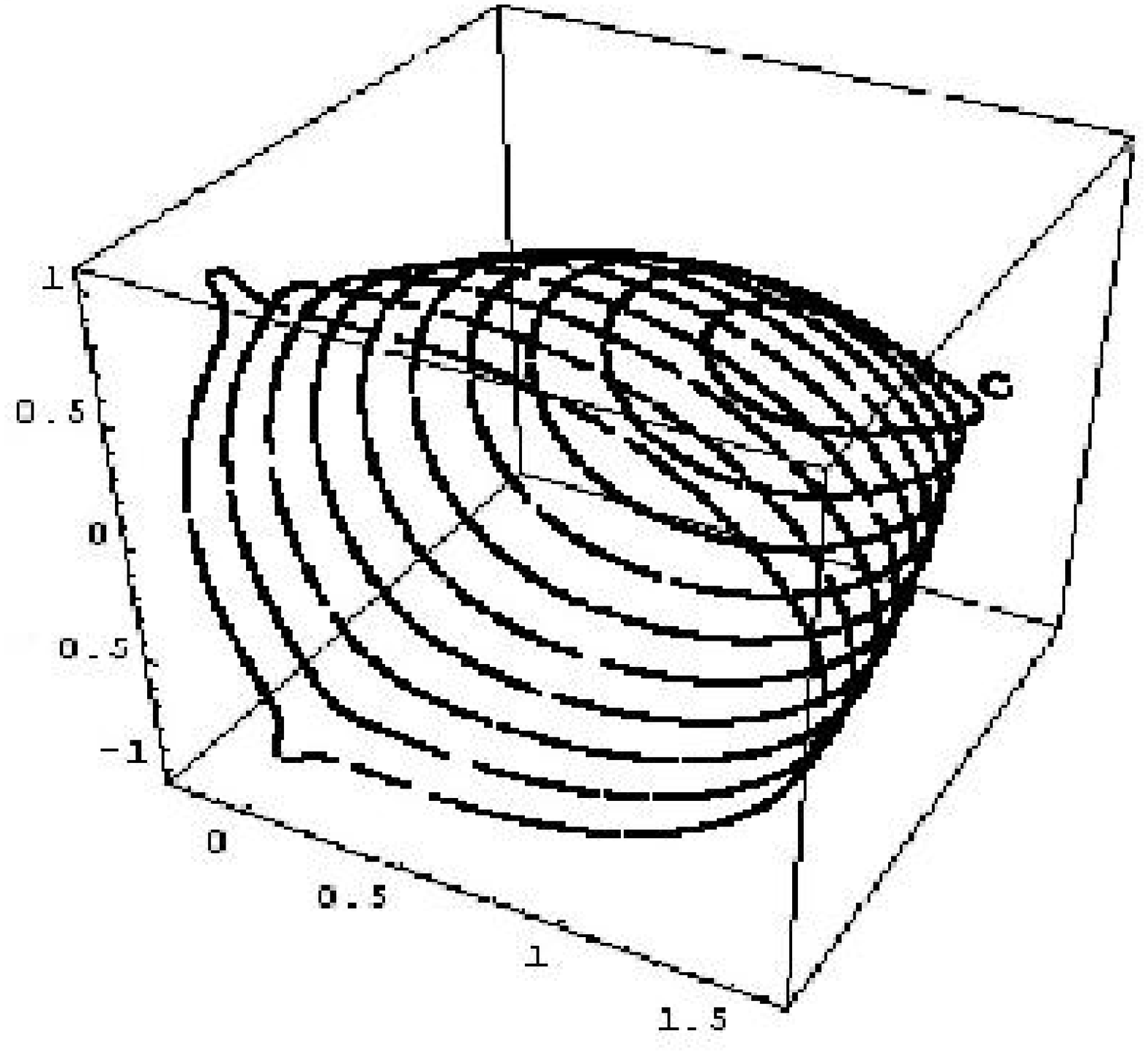}}
  \hspace{30mm}
\centering\subfigure[~Inner equi-$\gamma$ curves]
 {\includegraphics[width=50mm,height=40mm]{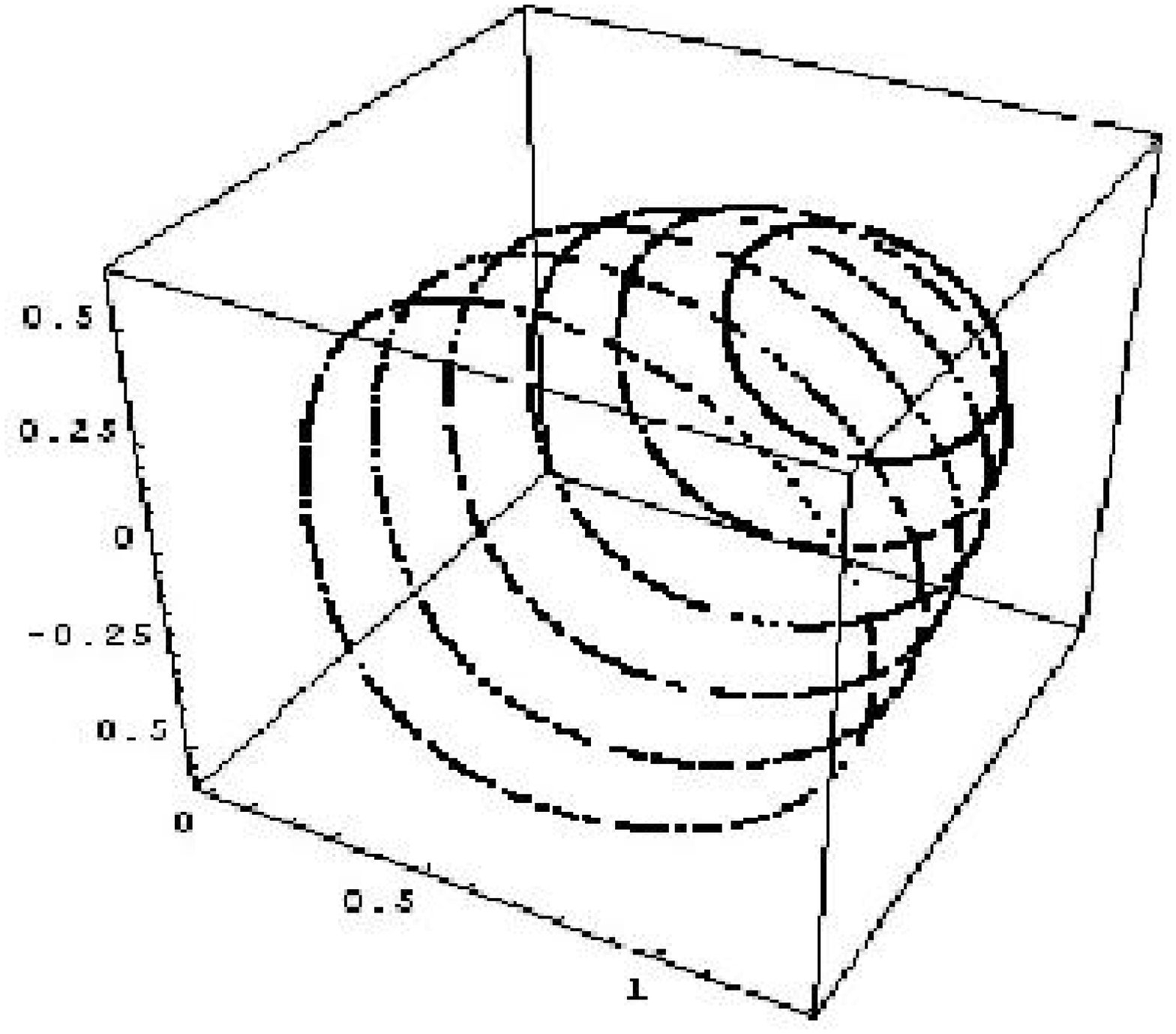}}
 \vspace{-2mm}
 \caption{\small{3-dimensional structures of tetrahedron shape with
 $a=0.5$
 }}
 \label{fig13}
\end{figure}

% (그림 14)
\begin{figure}[]
\centering \subfigure[~Outer equi-$\gamma$ curves]
{\includegraphics[width=50mm,height=40mm]{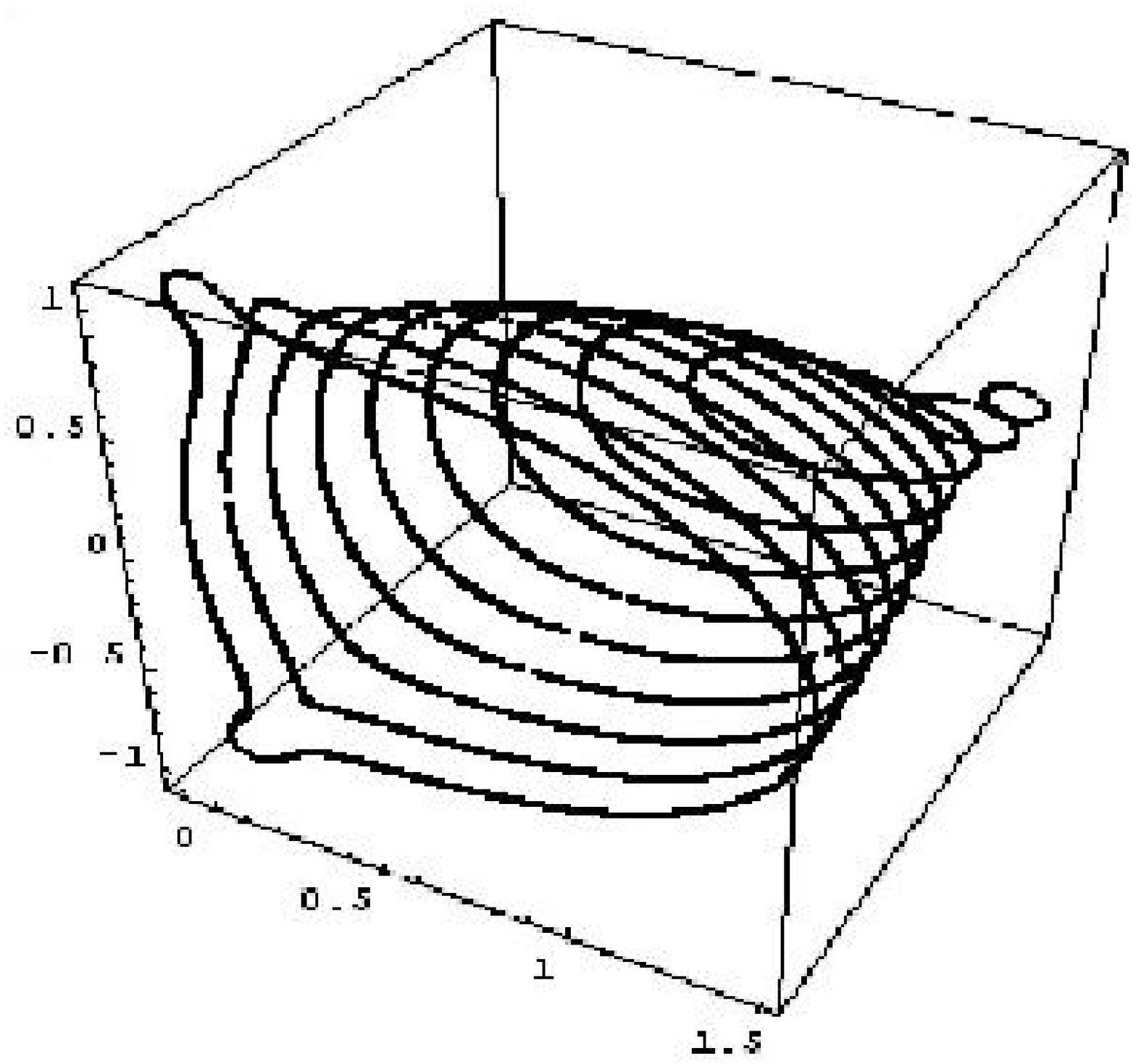}}
 \hspace{30mm}
\centering\subfigure[~Inner equi-$\gamma$ curves]
 {\includegraphics[width=50mm,height=40mm]{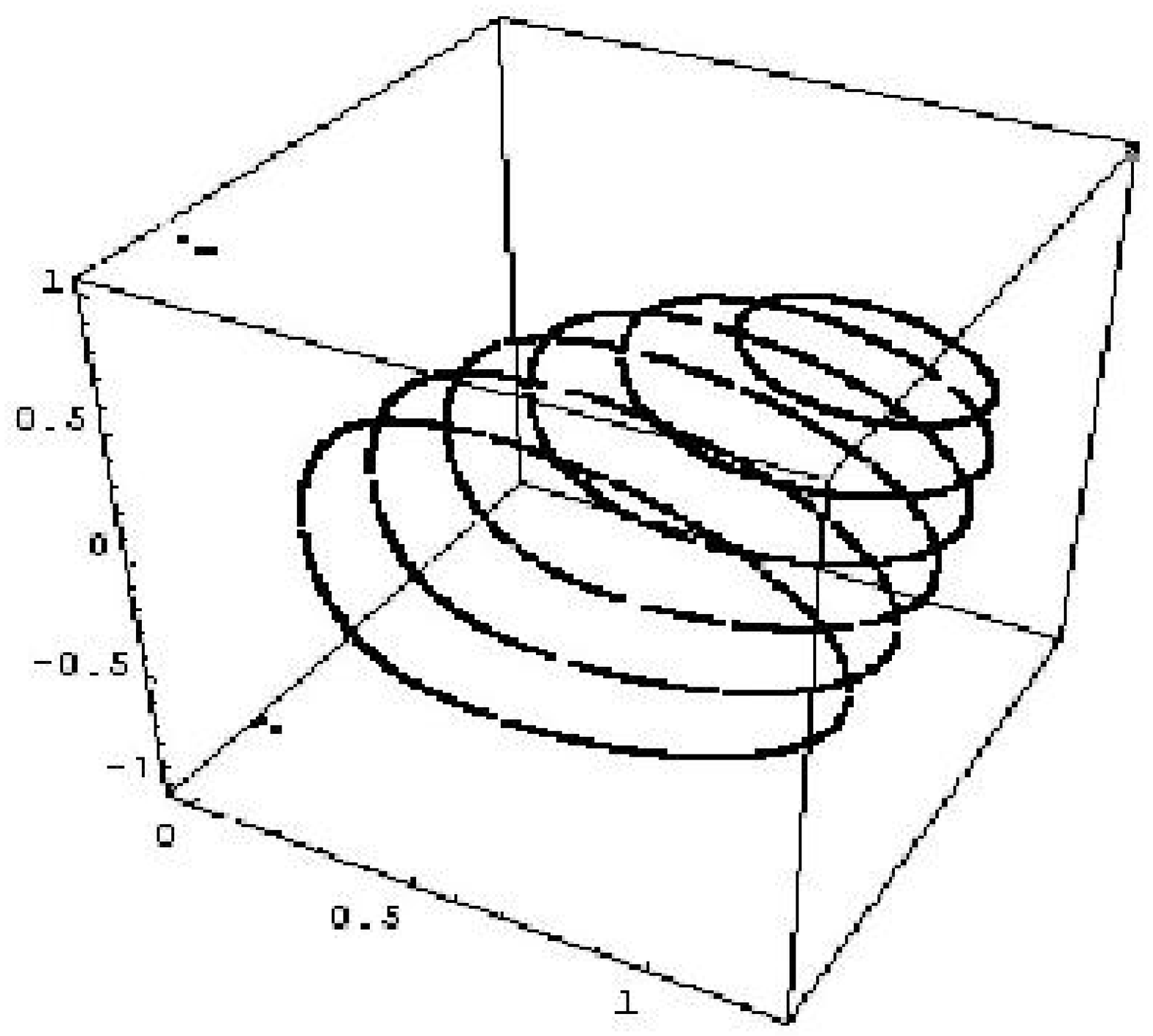}}
 \vspace{-2mm}
 \caption{\small{3-dimensional structures with $a=1.0$}}
 \label{fig14}
\end{figure}

%(그림 15)
\begin{figure}[]
\centering \subfigure[~Outer equi-$\gamma$ curves]
 {\includegraphics[width=50mm,height=38mm]{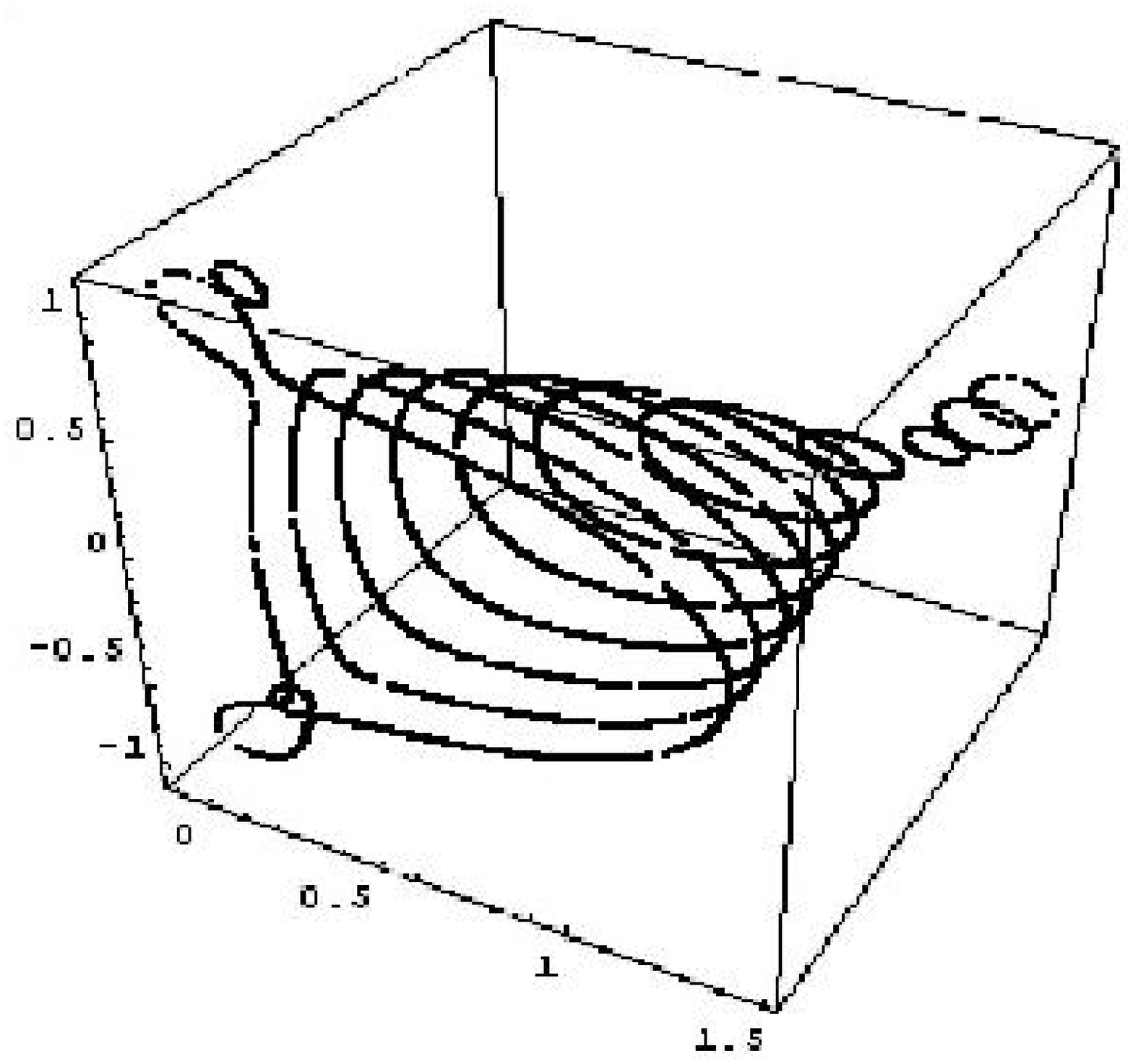}}
 \hspace{30mm}
\centering\subfigure[~Inner equi-$\gamma$ curves]
 {\includegraphics[width=50mm,height=38mm]{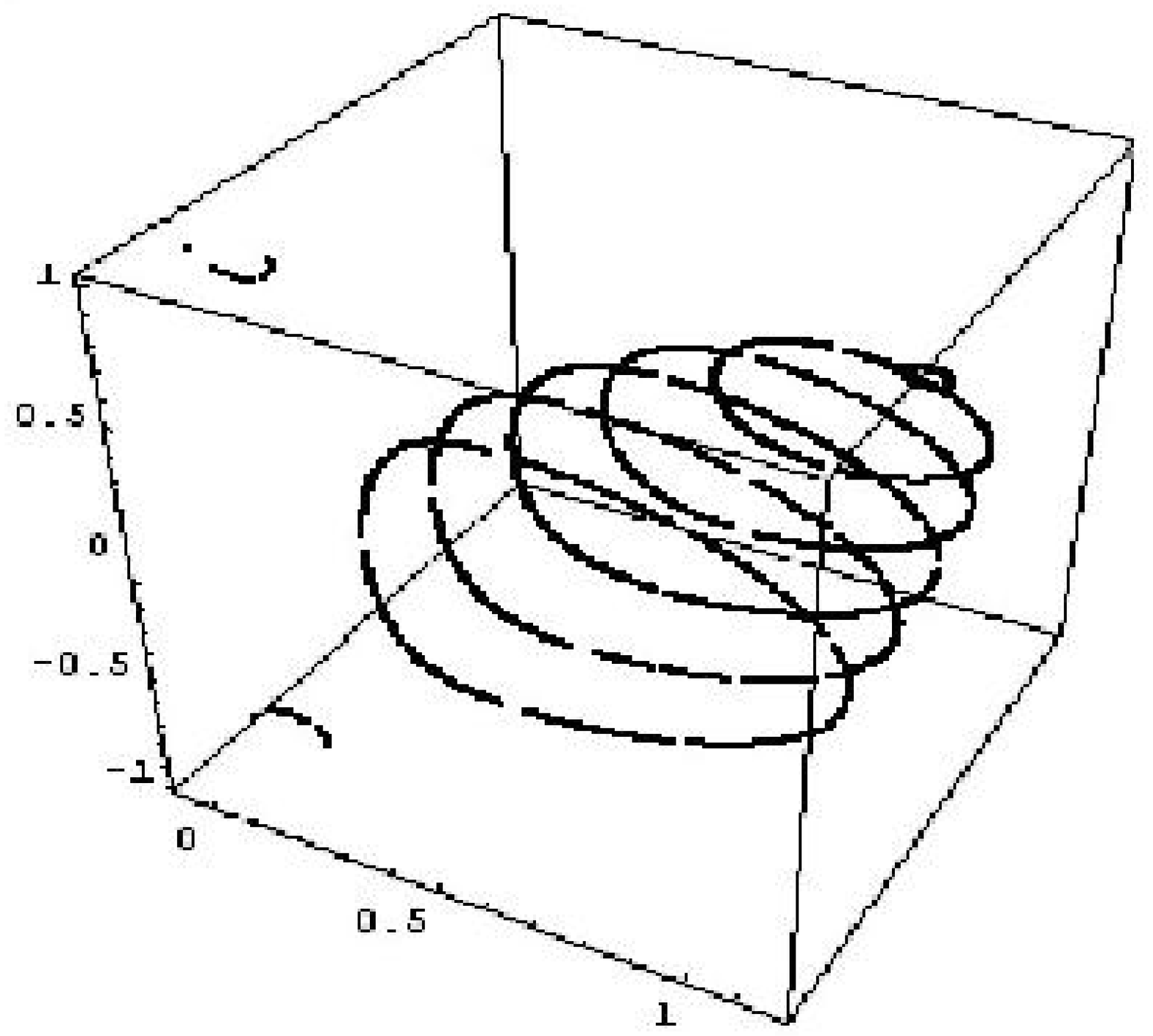}}
 \vspace{-2mm}
 \caption{\small{3-dimensional structures with $a=2.0$}}
 \label{fig15}
\end{figure}
\vspace{6cm}

%4절
%

\section{Discussions}
\label{four}
 In this paper, we have given a systematic formulation of flux-tubes
which can be applied to estimate the gluonic structures of hadrons
quantitatively. We can classify the flux-tubes and construct
topological spaces with simple assumptions. For the constructed
topological spaces, we can define connection amplitude which can be
used to deduce flux-tube overlap functions for various
configurations of boundary quarks. We have presented our results for
tetraquarks which are now of special interests after the discovery
of $X(3872)$. Since we do not know the configurations of quarks, we
have carried out the calculations for two different situations.
These different situations can be compared if the total gluon
densities are obtained by integration, however, the relative sizes
between the two situations are not known.\par
 The extension to pentaquarks or hexaquarks is immediate in our formalism.
But the most critical loophole lies in the process of fixing the
positions of boundary quarks. This problem is more evident if we
consider the simpler system of baryons. For example, in case of a
proton, we still do not know the relative positions and motions of
valence quarks\cite{R12}. In order to analyze the probed results, we
have to introduce various kinds of generalized parton
distributions\cite{R13} which are the main subjects discussed in
recent SIR Workshop at J-lab. Up to the present knowledge, the
quarks appear to have orbital angular momentum which is in some way
related to the gluonic motions or structures in proton. These
relations have to be studied more in the future. The spatial
dependences of gluonic structures may be closely related to the
concept of impact parameter dependent distributions\cite{R14}.\par
 In fact, we have defined the flux-tube overlap function $\gamma$ as a
product of the initial and the final connection amplitudes $A_{i}$
and $A_{f}$. For fixed quark positions, $A_{i}$ becomes a constant
and therefore has no effects on the form of $\gamma$. Then there
could be the questions about the meaning of $A_{i}$ and the
interpretation of $\gamma$ as a product of two amplitudes. One
possible answer can be sought by considering the initial motions of
quarks or changes of relative distances between quarks. Since the
quark distribution amplitudes are related to these variations,
$A_{i}$ may be used to represent the quark distributions. In our
calculations, only the contributions of $A_{f}$ have been considered
and it turns out that the main gluonic contributions come from the
central region surrounded by four quarks. This is the same for the
two cases considered, i.e., rectangular shape and tetrahedron shape.
Of course, there appear singular behaviors around boundary quarks,
which have to be cutoff at some point where perturbative
calculations can be applied.\par
 As for the state $X(3872)$, there exist many models such as $c\bar{c}$\cite{R15}, hybrid
charmonium\cite{R16}, diquark-antidiquark\cite{R17},
glueball\cite{R18}, cusp at $D^0\bar{D}^0$ threshold\cite{R19}, and
$D^{*0}\bar{D}^0$ molecule\cite{R20}. Of these possibilities, models
of diquark-antidiquark and  $D^{*0}\bar{D}^0$ molecule correspond to
tetraquarks. However, the more compact tetraquark state will be a
state with equal footings of the four quarks. In our case,
tetrahedron shape can be taken to be in that state. In contrast,
rectangular configuration will correspond to diquark-antidiquark or
$D^{*0}\bar{D}^0$ molecule state. The discriminations of these
models based on more data and more theoretical works have to be done
in the future.

\section*{Acknowledgements}
 This work was supported in part by the Fund of Chonbuk National
University. Jong B. Choi thanks Prof. Ji and the Department of Physics in
North Carolina State University for their invitation and hospitality during
his stay.

%참고문헌
%


\begin{thebibliography}{99}
\bibitem{R1} E. Eichten and F. L. Feinberg, Phys. Rev. {\bf D23}, 2724 (1981);
D. Gromes, Phys. Rep. {\bf 200}, 186 (1991).
\bibitem{R2} J. Carlson, J. Kogut, and V. R. Pandharipande, Phys. Rev. {\bf D27}, 233 (1983);
N. Isgur and J. Paton, {\it{ibid}}. {\bf D31}, 2910 (1985).
\bibitem{R3} R. Kokoski and N. Isgur, Phys. Rev. {\bf D35}, 907 (1987).
\bibitem{R4} P. Geiger and N. Isgur, Phys. Rev. {\bf D44}, 799 (1991).
\bibitem{R5} J. B. Choi and S. U. Park, J. Korean Phys. Soc. {\bf 24}, 263 (1991).
\bibitem{R6} J. B. Choi and W. J. Kim, J. Korean Phys. Soc. {\bf 25}, 477 (1992).
\bibitem{R7} J. B. Choi and H. Y. Choi, AIP {\bf CP494}, 353 (1999).
\bibitem{R8} S. K. Choi {\it{et al}}. [Belle Collaboration], Phys. Rev. Lett. {\bf 91}, 262001 (2003);
D. Acosta {\it{et al}}. [CDF II Collaboration], {\it{ibid}}. {\bf
93}, 072001 (2004); V. M. Abazov {\it{et al}}. [D0 Collaboration],
{\it{ibid}}.{\bf 93}, 162002 (2004); B. Aubert {\it{et al}}. [BABAR
Collaboration] Phys. Rev. {\bf D71}, 071103 (2005).
\bibitem{R9} T. Nakano {\it{et al}}., Phys. Rev. Lett. {\bf 91}, 012002 (2003).
\bibitem{R10} D. Diakonov, V. Petrov, and M. Polyakov, Z. Phys. A{\bf 359}, 305 (1997).
\bibitem{R11} R. Jaffe, Phys. Rev. Lett. {\bf 38}, 195 (1977).
\bibitem{R12} X. Ji, Phys. Rev. Lett. {\bf 91}, 062001 (2003); A. V. Belitsky, X. Ji,
and F. Yuan, Phys. Rev. {\bf D69}, 074014 (2004).
\bibitem{R13} D. M$\ddot{\rm{u}}$ller, D. Robaschik, B. Geyer, F.-M. Dittes, and J. Horejsi,
Fortschr. Phys. {\bf 42}, 101 (1994); X. Ji, Phys. Rev. Lett. {\bf
78}, 610 (1997); Phys. Rev. {\bf D55}, 7114 (1997); A. V.
Radyushkin, {\it{ibid}}. {\bf 56}, 5524 (1997).
\bibitem{R14} M. Burkardt, Phys. Rev. {\bf D62}, 071503 (2000).
\bibitem{R15} T. Barnes and S. Godfrey, Phys. Rev. {\bf D69}, 054008 (2004); E. Eichten,
K. Lane, and C. Quigg, {\it{ibid}}. {\bf 69}, 094019 (2004).
\bibitem{R16} F. E. Close and S. Godfrey, Phys. Lett. {\bf B574}, 210 (2003); B. A. Li, {\it{ibid}}.
{\bf 605}, 306 (2005).
\bibitem{R17} L. Maiani, F. Piccinini, A. D. Polosa, and V. Riquer, Phys. Rev. {\bf D71}, 014028
(2005).
\bibitem{R18} K. K. Seth, Phys. Lett. {\bf B612}, 1 (2005).
\bibitem{R19} D. V. Bugg, Phys. Lett. {\bf B598}, 8 (2004).
\bibitem{R20} N. A. T$\ddot{\rm{o}}$rnqvist, Phys. Lett. {\bf B590}, 209 (2004); C. Y. Wong, Phys. Rev. {\bf C69},
055202 (2004); F. E. Close and P. R. Page, Phys. Lett. {\bf B578},
119 (2004); E. S. Swanson, {\it{ibid}}. {\bf 588}, 189 (2004); E.
Braaten and M. Kusunoki, Phys. Rev. {\bf D69}, 074005 (2004).
\end{thebibliography}
\end{document}